\documentclass[conference]{IEEEtran}
\IEEEoverridecommandlockouts
\usepackage{cite}
\usepackage{amsmath,amssymb,amsfonts}
\usepackage{algorithmic}
\usepackage{graphicx}
\usepackage{textcomp}
\usepackage{xcolor}
\usepackage{times}
\usepackage{soul}
\usepackage{url}
\usepackage[hidelinks]{hyperref}
\usepackage{subcaption}  
\usepackage{array} 
\usepackage{caption} 
\usepackage{float}
\usepackage[linesnumbered,ruled,vlined,commentsnumbered,noend]{algorithm2e}
\usepackage{amsmath}
\usepackage{comment}
\usepackage{siunitx}
\newcommand{\system}{AutoHet\xspace}

\usepackage{cite}
\usepackage{amsmath,amssymb,amsfonts}
\usepackage{algorithmic}
\usepackage{graphicx}
\usepackage{textcomp}
\usepackage{xcolor}

\def\BibTeX{{\rm B\kern-.05em{\sc i\kern-.025em b}\kern-.08em
    T\kern-.1667em\lower.7ex\hbox{E}\kern-.125emX}}
\begin{document}

\title{Diving into 3D Parallelism with Heterogeneous Spot Instance GPUs: Design and Implications}

\author{Yuxiao Wang, Yuedong Xu, Qingyang Duan and Yuxuan Liu, Lei Jiao, Yinghao Yu, Jun Wu
\thanks{Yuedong Xu is with College of Computer Science and Artificial Intelligence, and Artificial Intelligence Innovation and Incubation Institute, Fudan University, Shanghai, China (e-mail: ydxu@fudan.edu.cn)}
}

\maketitle

\begin{abstract}
The rapid growth of large language models (LLMs) and the continuous release of new GPU products have significantly increased the demand for distributed training across heterogeneous GPU environments. In this paper, we present a comprehensive analysis of the challenges involved in implementing 3D parallelism in such environments, addressing critical issues such as the need for symmetric tensor parallelism, efficient gradient synchronization in asymmetric pipeline parallelism, and the trade-offs between memory utilization and computational efficiency. Building upon these insights, we introduce \system, a novel system that automatically identifies the optimal parallelism plan for distributed training on heterogeneous GPUs. \system supports asymmetric 3D parallelism structures and facilitates fine-grained workload distribution. We propose a theoretical model that frames the device grouping and load balancing as an optimization problem to minimize per-iteration training time, thus effectively balancing computing power and memory usage across GPUs with diverse capabilities. To enable elastic training upon spot instance preemption, \system presents an efficient recovery strategy that prioritizes to retrieve training states from local nodes, and only downloads the missing checkpoints from the cloud storage. Our extensive evaluation, conducted on three large-scale models and utilizing combinations of three different GPU types, demonstrates that \system outperforms existing DNN training systems, achieving up to a 1.79$\times$ speedup in training throughput compared with Megatron-LM and Whale, and a 4.38$\times$ speedup of recovery speed compared to a spot instance baseline. 
\end{abstract}


\section{Introduction}

\begin{figure}[t]
	\centering
	\includegraphics[width=\linewidth]{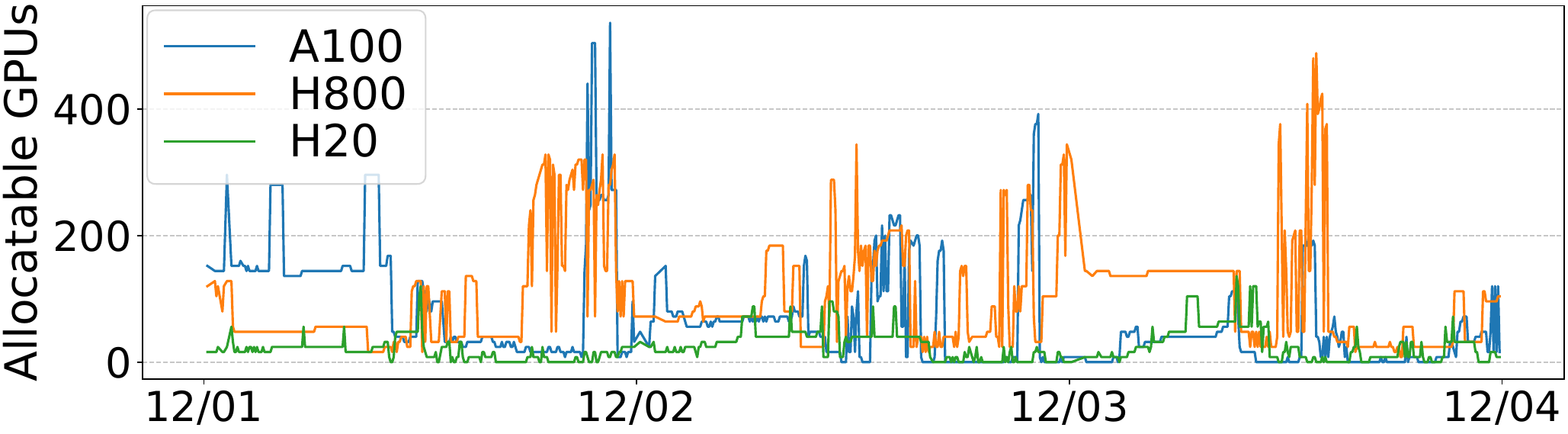}
	\caption{Allocable GPU spot instances over time.}
	\label{fig:available GPUs}
\vspace{-0.5cm}
\end{figure}

The ``arms race'' in large language models (LLM) like GPT-3~\cite{gpt3}, Gemini~\cite{team2024gemini}, LLaMA~\cite{touvron2023llama} etc., has driven the rapid advancement of GPUs, with their computing power and storage capacity doubling every two to three years~\cite{nvidia_blackwell_architecture,nvidia_cudagpus}. Due to the relatively long lifespan of GPUs, different types of GPU machines are often mixed within a computing cluster~\cite{hetecluster,weng2022mlaas,jiang2017heterogeneity}. It is foreseeable that the GPU types in future clusters will be even more diversified. An intriguing phenomenon in today's production clusters is the fluctuation in GPU availability, especially in the spot instance scenarios~\cite{amazonec2,microsoftazure}, where resources are not always abundant. The high demand for training further exacerbates supply shortages, leading to unacceptable delays for users seeking homogeneous GPUs. Figure~\ref{fig:available GPUs} illustrates the variation in allocable GPU availability over three days in our cluster. At a given snapshot, homogeneous GPUs may be insufficient for large-scale model training, highlighting the need for advanced frameworks that can efficiently utilize diverse GPU resources to optimize training performance.

To effectively pre-train LLMs, existing training frameworks~\cite{2019megatron,2021megatron,2020deepspeed} have made significant advances in optimizing 3D parallelism, which combines data parallelism (DP)~\cite{pytorch,rajbhandari2020zero}, tensor parallelism (TP)~\cite{2019megatron,2021tp,shazeer2018mesh}, and pipeline parallelism (PP)~\cite{pipedream2bw,narayanan2019pipedream,huang2019gpipe} to distribute workloads across GPU clusters. While 3D parallelism perform well in homogeneous environments, it typically assume uniformity in GPU resources. This assumption facilitates the coordination of parallel training strategies but becomes a critical limitation in heterogeneous environments. Recent efforts have been devoted to enhancing the performance of heterogeneous GPU training ~\cite{chen2020semi,ding2021hetseq,duan2022hph,jia2022whale,park2020hetpipe,song2020accpar,zhou2023abs}, while most of them are constrained to specific parallelism dimensions, such as adjusting batch-size within DP~\cite{zhou2023abs}, only TP~\cite{song2020accpar}, or only DP and PP~\cite{duan2022hph,park2020hetpipe}. Whale~\cite{jia2022whale} introduces a hardware-aware load balance algorithm designed to distribute workloads within intra-parallelism. SDPipe~\cite{miao2023sdpipe} propose 
a semi-decentralized training
system specialized for pipeline-parallel training in dynamic heterogeneity environments. HPH~\cite{duan2022hph} and HetPipe~\cite{park2020hetpipe} concentrate on layer load balancing under DP and PP, excluding TP from their scope. Metis \cite{um2024metis} presents an efficient algorithm to prune the large search spaces and balance loads with heterogeneity-awareness. This algorithm is built on a few observations that the sizes of differnt PP stages do not vary significantly, and increasing DP is more beneficial than TP. 


In this paper, we aim to address three main challenges that existing training systems do not effectively tackle. 
First, the current 3D parallelism structures and methodologies are overly simplistic in system design and implementation, often leading to sub-optimal parallel strategies. Existing training systems~\cite{2019megatron,2020deepspeed,jia2022whale,2021megatron} predominantly rely on symmetric GPU distribution and parallel structures, where 
each parallel group is required to exhibit the same degree of parallelism. This symmetry constraint severely limits the exploration space for optimal parallel strategies in environments with diversified GPU computational and memory capacities. 
Second, the inherent asymmetry introduced by the first challenge significantly complicates load balancing. Without efficient load balancing, more powerful GPUs may remain underutilized while less capable GPUs become bottlenecks, leading to sub-optimal overall training performance. Third, rapid training recovery in the event of GPU spot instance preemption is rarely studied, despite that the dynamic availability of different GPU types is a key reason for adopting heterogeneous GPUs in training.




To overcome these challenges, we present \system, an automated 3D parallel training system designed for heterogeneous GPU environments. First, we perform a few simple but enlightening experiments to reveal the important properties of training with heterogeneous GPUs. In particular, TP cannot be asymmetric because of its high matrix transpose overhead before \textit{All-Reduce} communication, while PP can be asymmetric across different DP groups. These observations allow us to eliminate arbitrary combinations of TP and PP in the subsequent planning of model training. Second, we present a two-stage decomposition approach to unravel the complexity of 3D parallel training with heterogeneous GPU types and communication links. At stage one, we formulate a nonlinear mixed-integer programming problem for planning 3D parallelism by maximizing the effective computing power, and obtain the optimal assignment of GPUs in DP groups. At stage two, we map each GPU to a certain PP stage of a DP group in which TP only operates on the GPUs connected via NVLinks, and perform the model partitioning on all the PP stages for computational load balancing. Third, we design an efficient migration strategy to resume training under varying parallelization plans. It prioritizes retrieving checkpoints locally and fetches only the missing ones from cloud storage, significantly reducing recovery time.

We evaluate \system's training efficiency across three GPU types (A100, H800, and H20) and three model architectures (BERT-Large~\cite{2019-bert}, LLaMA, and GPT-3) on a platform equipped with 24 GPUs in total. AutoHet achieves 1.38$\times$ speedup over Megatron-LM for BERT-Large, 1.53/1.27$\times$ over Megatron-LM/Whale for GPT-3, and up to 1.79/1.51$\times$ under non-uniform GPU distributions for LLaMA.  In scalability analysis with simulated configurations up to 64 GPUs, planning overhead ranges from 1.23-159.12 seconds, with profiling time of 11.9-15.4 minutes, nearly ten times faster than Alpa. AutoHet's migration strategy delivers up to 4.38$\times$  faster recovery compared to Varuna~\cite{varuna} through optimized checkpoint management.



\section{Background and Observations}
\label{sec:background}


\subsection{DNN Training Parallelism}
Existing LLM parallel training techniques typically include data parallelism (DP), tensor parallelism (TP), and pipeline parallelism (PP). 

\textbf{\textit{1) Data parallelism}} involves distributing a partitioned dataset across various training nodes~\cite{dean2012large}. Each node possesses a full replica of the model, conducts training on its segment of data, and subsequently contributes to gradient exchange and parameter synchronization so as to construct a global model. This iterative process persists until the model attains convergence.

\textbf{\textit{2) Tensor parallelism}} refers to a method where a neural network tensor is segmented into numerous equal-sized blocks along a specific dimension~\cite{2019megatron,2021megatron}. Consequently, the operations on these tensor fragments are executed in parallel across multiple nodes, enhancing computational efficiency by distributing the workload.

\textbf{\textit{3) Pipeline parallelism}} refers to an approach wherein a model is segmented into various stages by layers, and each segment is allocated to a different computational node~\cite{huang2019gpipe,narayanan2019pipedream,pipedream2bw}. The intermediate result of a preceding node is transmitted to its succeeding node as the input. 

\textbf{\textit{4) 3D parallelism. }}With the ever-increasing model size, using only one type of parallelization is inefficient. The aforementioned techniques are usually combined as a method we call 3D parallelism. A lot of efforts have been devoted to orchestrating 3D parallelism to improve the utilization of GPU streaming cores~\cite{2019megatron,2021megatron,bekman2022bloom,2020deepspeed}. 

In what follows, we conduct an in-depth exploration of multidimensional parallelism with heterogeneous GPUs, presenting key insights toward the design of automatic parallelism. 

\subsection{Symmetric Tensor Parallelism with Hetero-GPUs} 
\label{subsec:symmetric tp}
\noindent\textbf{Observation 1}: \emph{Tensor parallelism partitioning needs to be symmetric across different DP chains. }

\begin{figure}[t]
        \centering
        \includegraphics[width=\linewidth]{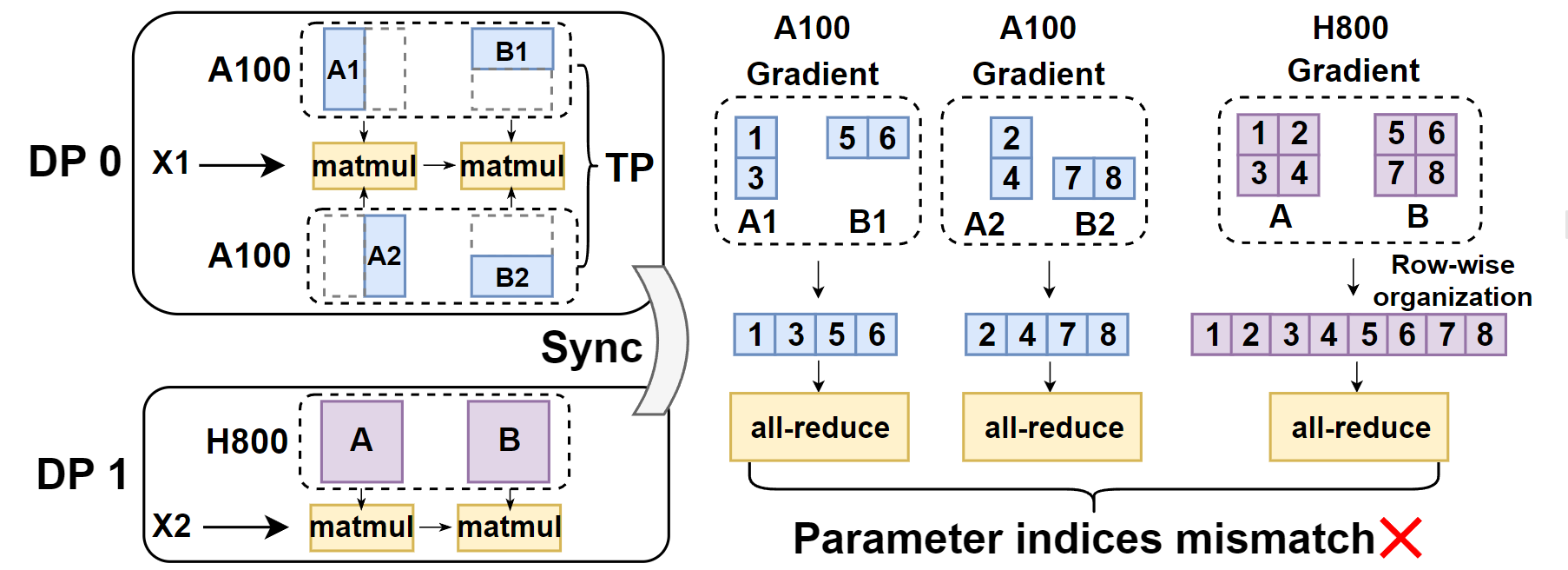}
        \caption{Illustration of asymmetric TP and gradient aggregation.}
        \label{fig:asy_tp}
\vspace{-0.5cm}
\end{figure}

We consider a hybrid TP and DP setting with one NVIDIA H800 GPU and two collocated A100 GPUs. Given the discrepancy of computing power in A100 and H800, and the high bandwidth NVLink interconnecting two A100 GPUs, it is reasonable to align two A100 GPUs for TP, and then let this TP group work with H800 GPU for DP (as shown in Figure~\ref{fig:asy_tp}). To synchronize the gradients inside this DP group, an AllReduce operation is executed in each iteration. 

The core of existing deep learning frameworks, e.g.,~\cite{2019megatron,2020deepspeed} , is the general matrix to matrix multiplication (GEMM). When a transformer block is partitioned on two GPUs, their GEMM process also changes. Taking a two-layer MLP tensor as an example that involves two parameter matrices $A$ and $B$. We segment $A$ vertically into $A_1$ and $A_2$, and segment $B$ horizontally into $B_1$ and $B_2$. The gradient computed on each GPU takes the same form as the parameter. Denote $g_A$ and $g_B$ as the gradients on the parameter matrices $A$ and $B$, respectively. 
Here, $g_{A_1}$ and $g_{B_1}$ are stored as two vectors $[1\;3]$ and $[5\;6]$ in the first A100 GPU, and $g_{A_2}$ and $g_{B_2}$ are stored as two vectors $[2\;4]$ and $[7\; 8]$ in the second A100 GPU. While in H800 GPU, $g_A$ and $g_B$ are stored as two vectors $[1\;2\;3\;4]$ and $[5\;6\;7\;8]$. When AllReduce is called, $g_{B_1}$ and $g_{B_2}$ from two A100 GPUs can be directly aggregated with $g_{B}$ from H800 GPU. However, $g_{A}$ must be transposed to match $g_{A_1}$ and $g_{A_2}$ for gradient aggregation, which incurs high computation complexity and requires additional temporary storage space.

\begin{figure}[t]
        \centering
        \includegraphics[width=\linewidth]{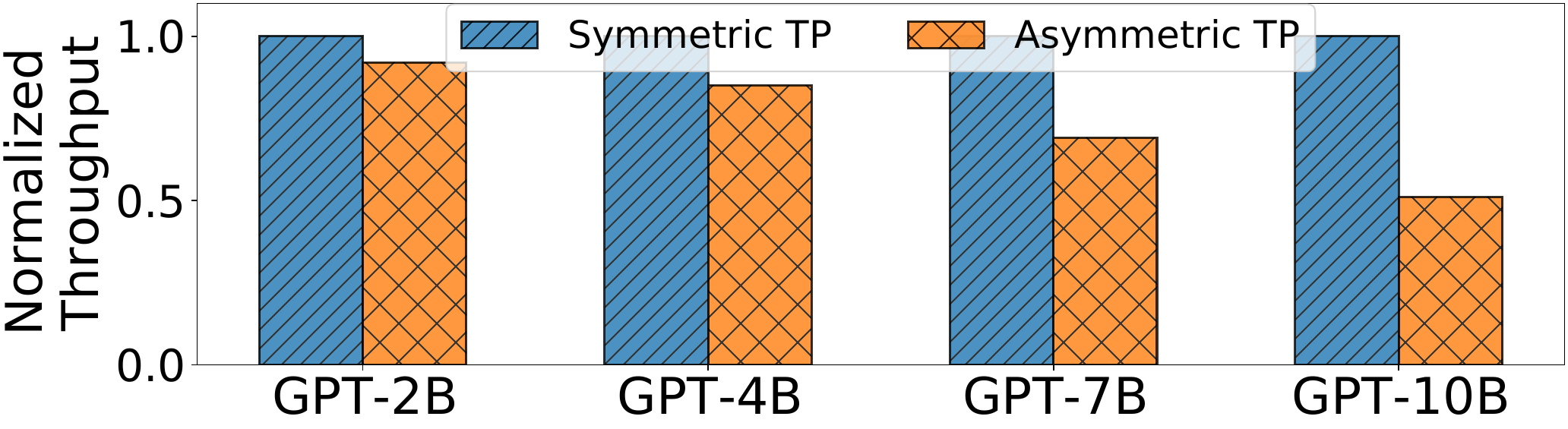}           
        \caption{Normalized training throughput with varying model sizes under different TP configurations. For 2B and 4B models, two configurations are compared: [A100×2, A100] vs. [A100, A100]; for 7B and 10B models, another two are compared: [A100×2, A100×2] vs. [A100×4, A100x2].
}
        \label{fig:asy_tp_overhead}
        \vspace{-0.5cm}
\end{figure}

We modify Megatron-LM in order to support the above-mentioned asymmetric TP. Figure~\ref{fig:asy_tp_overhead} shows the performance overhead of introducing asymmetric TP for different model sizes. Asymmetric setups are created by only adding GPUs to symmetric configurations for asymmetric TP, while other settings are unaltered so as to ensure the identical throughput when such transpose overhead does not exist. 
The degradation of training throughput ranges from 8\% to 49\% for the models up to 10B parameters and becomes even worse with larger models. Without changing the existing design of TP, we conclude that the tensor parallelism must be symmetric across different DP chains.
\subsection{Asymmetric Pipeline Parallelism with Hetero-GPUs}
\label{subsec:asy_pp} 
\noindent\textbf{Observation 2}: \emph{Asymmetric pipeline parallelism demands layer-wise gradient synchronization.}

\begin{figure}[t]
        \centering
        \includegraphics[width=\linewidth]{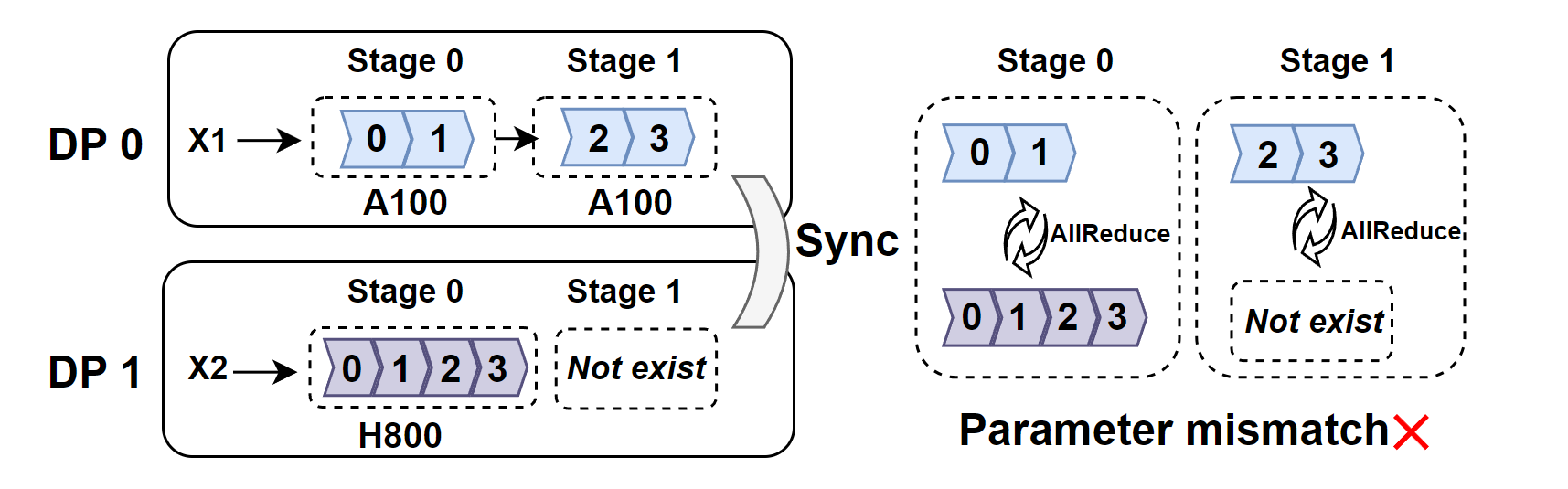}
        \caption{Illustration of asymmetric PP and gradient aggregation.}
        \label{fig:asy_pp}
        \vspace{-0.5cm}
\end{figure}

When heterogeneous GPUs are assigned to multiple pipelines, the term ``pipeline stage'' becomes inconsistent in different data parallel groups. In our scenario, two A100 GPUs are concatenated in a pipeline that works with an H800 GPU for data parallelism. The LLM to be trained consists of four transformer layers, where each A100 GPU stores two consecutive layers, and the H800 GPU stores the entire model, as shown in Figure~\ref{fig:asy_pp}. After backpropagation is completed, an AllReduce operation is performed to synchronize the gradients across the data parallel groups. 

In the frameworks optimized for homogeneous GPU training \cite{2019megatron,2021megatron,2020deepspeed}, all GPUs within the same stage synchronize their gradients, typically communicating in a ring topology where each GPU exchanges data with a single neighboring GPU. 
The first DP group contains two stages, whereas the second one contains a single stage. H800 GPU transmits layer-4 and layer-3 gradients to the second-stage A100 GPU, and layer-2 and layer-1 gradients to the first-stage A100 GPU. The both A100 GPUs also transmit all their gradients to H800 GPU. In light of this misalignment of pipeline stages among different DP groups, the AllReduce primitive must be adapted to accommodate this form of hybrid parallelism. Intuitively, the Ring AllReduce topology bifurcates if the gradient of a GPU is taken as a unit of data. When the number of DP groups increases, the AllReduce topology becomes extraordinarily complicated. One possible solution is to execute the AllReduce process at the granularity of individual layers, with each layer utilizing a different ring. 

\subsection{Complicated Load Balancing}
\label{subsec:memoryfilling}

\noindent\textbf{Observation 3}: \emph{A dilemma exists between GPU memory utilization and training efficiency on heterogeneous GPUs.} 

In homogeneous training, TP and PP are devised to conquer the GPU memory limit, and fully utilizing GPU memory is almost equivalent to maximizing the GPU computing power. However, this tenet does not hold in heterogeneous GPU training. Let us use a toy example to illustrate memory filling in a pipeline consisting of two A100 and two H800 GPUs. Recall that the actual computing power of H800 is twice that of A100 in our setting. Considering two memory filling methods, \emph{equal model partitioning} and \emph{proportional model partitioning}. In the former, the computing load of each GPU is nearly identical, but their durations of forward and backward passes differ considerably. 
The ratio of overall idle GPU time amounts to $75\%$. In fact, the the wasted computing power of an A100 GPU amounts to $\frac{2}{3}$, and that of an H800 GPU amounts to $\frac{5}{6}$, which means that equally partitioning the model is computationally inefficient. In the latter, the DNN layers are proportionally assigned to different GPUs according to their relative computing powers, and the fastest GPUs are filled with as many layers as possible. In this situation, a low-quality GPU (e.g., A100) is assigned with fewer layers, but its memory size is usually more than half of the high-quality one (e.g., H800), so its memory is underutilized. It can be readily concluded that balancing computational workloads across heterogeneous GPUs can accelerate the model training. However, it remains essential to consider the utilization of available memory resources.

\section{\system 3D Parallelism}
\label{sec:systemdesign}


\subsection{Design Rationale}
\label{subsec:modeloverview}

\begin{figure*}[htbp]
  \centering
  \includegraphics[width=0.80\textwidth]{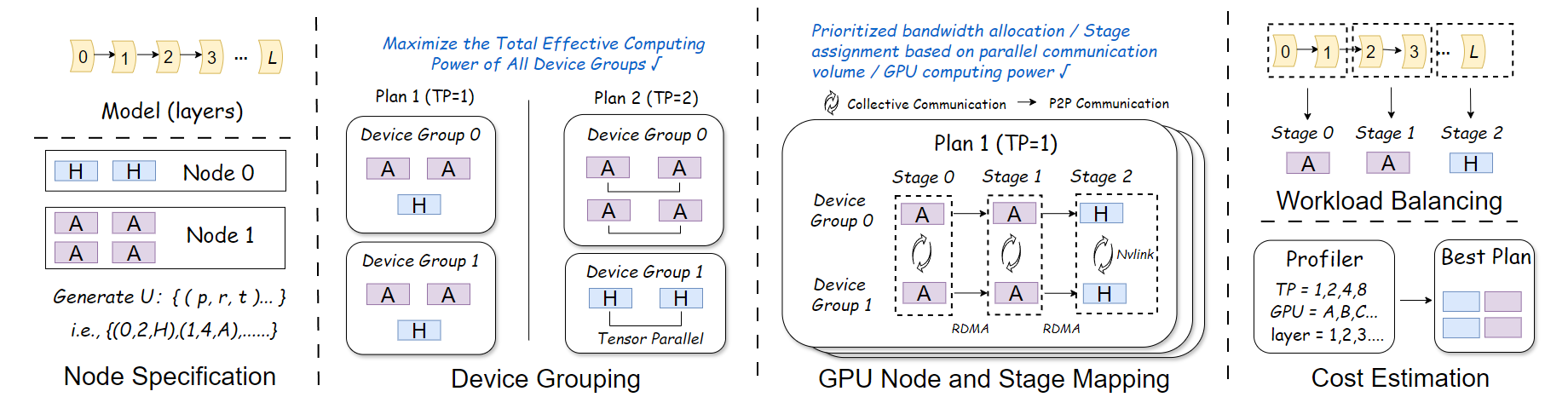}
  \caption{\system’s Execution Procedure.}
  \vspace{-0.2cm}
  \label{fig:alg overview}
  \vspace{-0.5cm}
\end{figure*}

We begin with a general cost model of 3D parallel training in which the widely used 1F1B pipeline scheduler ~\cite{narayanan2019pipedream} is adopted. Denote by $T^{\ast}$ the per-iteration training time:
\begin{small}
\begin{align}
T^{\ast} &= \min \left( \max_{j \in D} \left\{ \sum_{i=1}^{P} t_{i}^{j} + (K - 1) \cdot \max_{c \in S} t_{c}^{j} \right\} + T_{sync} \right )
\end{align}
\end{small}
where $P$ denotes the total number of stages, $K$ is the number of micro-batches, $t_{i}^{j}$ represents the computation time for the forward pass and backward pass of the $i$-th stage within the $j$-th data parallel group (including TP and PP communications), and $T_{sync}$ is the time required for gradient synchronization, calculated by dividing the data volume for model synchronization by the lowest communication bandwidth among the data parallel groups. 

This cost model abstracts away the underlying complexity of 3D parallelism with heterogeneous GPUs and diverse communication links. When all the GPU are homogeneous, we only need to specify the number of GPUs assigned to TP, PP and DP groups, and the model partitioning is symmetric. However, in a heterogeneous scenario, the 
optimal parallelization strategy involves selecting appropriate DP, TP, and PP groupings for each GPU while taking into account workload distribution, GPU resource constraints, and host server associations, which results in significant complexity. To address this challenge, we decompose the problem into two stages: i) computing power maximization and ii) GPU mapping and model partitioning. 


The workflow of \system is illustrated in Figure~\ref{fig:alg overview}.
At the beginning, it generates multiple balanced 3D parallel device grouping plans based on the input model and node specifications, temporarily disregarding their physical locations. GPU placement optimization follows within each plan, determining an efficient mapping of nodes and pipeline stages. The model is then partitioned across the pipeline stages for load balancing, and the best plan is selected through lightweight profiling and cost estimation. 



\subsection{Effective Computing Power Maximization}
\label{sec:3D Parallel Planning Algorithm}

Our goal is to assign heterogeneous GPUs for 3D parallelism by considering the balancing of computational load on different data parallel groups. 
A device group is defined as a collection of an arbitrary number of homogeneous or heterogeneous GPUs that collectively handle a complete model.
To ensure consistent model accuracy and synchronous gradient aggregation, the computing power between device groups must be kept as balanced as possible without modifying the batch size. We model this as a nonlinear integer optimization problem to determine the optimal device grouping plan.

\noindent\textbf{\textit{Node specification.}} \system supports heterogeneous GPU clusters, allowing for variations in both the number and type of GPUs across different nodes. The heterogeneous GPU configuration $S$ is formulated as a set of 3-tuples, such as \{(0, 8, A100), (1, 4, H800), …\}, indicating the presence of eight A100 GPUs on node 0, four H800 GPUs on node 1, and so forth. 

\noindent\textbf{\textit{Problem formulation.}} The total number of GPUs is denoted by $N$. The variable \(x_{i,j} \in  \left \{ 0,1 \right \}\) represents whether the \(i^{th}\) GPU is assigned to the \(j^{th}\) DP group. Similarly, $y_{j} \in  \left \{ 0,1 \right \}$ indicates the presence of at least one GPU in the $j^{th}$ DP group. Since each DP group must contain at least one GPU, the number of DP groups cannot exceed $N$. The memory capacity of each GPU is denoted as $m_{i}$ for the $i^{th}$ GPU. Because TP need to be symmetric (as detailed in Section~\ref{subsec:symmetric tp}), the modeling can be simplified to focus on the combination of DP and PP. We introduces a new metric called \textit{effective computing power}, symbolized by \(G_{j}\). This metric quantifies the actual computing power of the \(j^{th}\) DP group.
\begin{small}
\begin{equation}
G_{j} =  {\textstyle \sum_{i=1}^{N}} g_{i}\cdot x_{i,j}\ \cdot\left (1-\rho _{j}\right )
\end{equation}
\end{small}where $g_{i}$ quantifies the computing power of the $i^{th}$ GPU, while $\rho _{j}$ denotes the pipeline bubble ratio in $j^{th}$ DP group. The problem formulation is defined in Equation (\ref{model}). The core idea is to maximize the product of the number of valid DP groups and the minimum \textit{effective computing power}, thereby minimizing the per-iteration time.

In the objective function, we sum over all variables $y_{j}$ to determine the total number of valid DP groups. The variable $z=\min_{j} \left \{ {G_{j}} \right \}$ is introduced to represent the minimum \textit{effective computing power} across all valid DP groups. 
 
\begin{small}
\begin{subequations}
    \begin{align}
        \underset{\left \{ x_{i,j} \right \}}{\text{Maximize}}  \quad & \sum_{j=1}^{N} y_{j} \cdot z \label{model} \\
        \text{Subject to:} \ \ \ & \sum_{i=1}^{N} m_{i} \cdot x_{i,j} + L \cdot (1 - y_{j}) \geq \text{MIN}_{\text{mem}}, \  \forall j;  \label{constraint:1} \\
                                 & G_{j} \cdot y_{j} + L \cdot (1 - y_{j}) \geq z, \quad \forall j; \label{constraint:2} \\
                                 & \frac{1}{L} \cdot \sum_{i=1}^{N} x_{i,j} \leq y_{i} \leq \sum_{i=1}^{N} x_{i,j}, \quad \forall j; \label{constraint:3} \\
                                 & \sum_{j=1}^{N} x_{i,j} = 1, \quad \forall i. \label{constraint:4}
    \end{align}
\end{subequations}
\vspace{-0.2cm}
\end{small}

Here, Constraint (\ref{constraint:1}) ensures that each DP group is equipped with sufficient memory for training. We profile the minimum required memory $\text{MIN}_{\text{mem}}$ during model training. Constraint (\ref{constraint:2}) limits the range of minimum effective computing power in all valid DP groups. $L$ is a sufficient large constant, introduced as an auxiliary variable to handle the special case when a DP group is empty. Constraint (\ref{constraint:3}) characterizes the value of $y_{j}$, while Constraint (\ref{constraint:4}) ensures that each GPU is assigned to exactly one DP group. Due to the reduced complexity, we compute the optimal solution using the math solver SCIP \cite{scip} directly.

\subsection{GPU Mapping and Model Partitioning}

Once the effective computation power is maximized and balanced across all DP groups, \system needs to materalize this allocation strategy, i.e. mapping each model partition on every GPU. To meet the communication and storage constraints, the following principles are developed. 


\noindent\textbf{\textit{GPU node mapping. }}The mapping of GPU nodes determines the allocation of communication bandwidth across different parallel dimensions. In this context, we consider two levels of communication bandwidth: high-speed intra-node communication via NVLink (e.g., 600 GB/s for A100 GPUs) and low-speed inter-node communication via RDMA(e.g., 400Gb/s). Bandwidth allocation is prioritized according to communication volume, with TP operations receiving the highest priority, followed by DP, and then PP. Specifically, we ensure that all communications between TP operations are routed through NVLink, with any remaining NVLink bandwidth allocated to DP groups to maximize its utilization.

\noindent\textbf{\textit{Pipeline stage mapping. }}We find that earlier stages require more memory to store forward activations, aligning with our observation of underutilized memory in lower-end GPUs (detailed in \S \ref{subsec:memoryfilling}). Futhermore, since computation and communication overlap in pipeline parallelism, communication efficiency is mainly influenced by the earlier stages. Lower-end GPUs handle smaller workloads and generate less communication, so AutoHet assigns these GPUs to the earlier pipeline stages where memory demand is high but the computational load is low.

Building on the aforementioned principles, AutoHet develops a heuristic algorithm for mapping GPUs to specific physical nodes and pipeline stages. The key idea is to assign GPUs with lower communication overhead and higher available memory to the earlier pipeline stages, while allocating bandwidth sequentially based on communication priorities. Notice that AutoHet initiates the process by identifying valid TP dimensions (Line 2), which require the number of GPUs per node to be an integer multiple of the TP dimension. During subsequent processing steps, GPUs assigned to a TP group are treated as a single entity. The alogrithm begins by sorting all GPU types in the $type\_set$ according to their computing power. It then iteratively selects the GPU type with the lowest computing power, checking two conditions: i), whether each DP group has an unassigned GPU of this type; and ii), whether there are available GPUs on the same node for all DP groups. If both conditions are met, the algorithm assigns the GPU type to the earliest unassigned pipeline stage and allocates the corresponding rank on the physical node. This process continues until NVLink communication requirements between DP groups can no longer be fulfilled. Any remaining pipeline stages and physical nodes are then assigned in sequence.

Load balancing in pipeline parallelism aims to distribute the workload efficiently across all stages to avoid bottlenecks and ensure optimal resource utilization. To address this, we introduces an optimization model for effective model partitioning. 

\noindent\textbf{\textit{Problem formulation.}} Let $P$ be the number of PP stages in a DP group. $p_i$ represents the pp stage where the $i^{th}$ gpu is located, $l_{i}$ denote the number of model layers allocated to the $i^{th}$ GPU. $N_{layers}$ represent the total number of model layers. The objective of the optimization problem is to allocate model layers across GPUs in a manner that best aligns with their respective computing power:
\begin{subequations}
\begin{align}
    \underset{\left \{ 1 \leq i \leq P \right \}}{\text{Minimize}} \ \ \quad & \max_{i} \left\{ \frac{g_{i}}{l_{i}} \right\} \label{eq1} \\
    \text{Subject to:} \quad & N_{layers} = \sum_{i=1}^{P} l_{i}; \label{eq2} \\
    & \text{MEM}_{\text{F}}(l_i) + \text{MEM}_{\text{V}}(l_i, p_i) \leq m_{i}, \quad \forall i.\label{eq3}
\end{align}
\end{subequations}
Here, Equation (\ref{eq2}) is a natural constraint of storing all the layers of the large language model. Constraint (\ref{eq3}) restricts the allocation of layers to each stage, ensuring that the number of layers assigned does not exceed the stage’s memory capacity. We profile both the fixed memory components $\text{MEM}_{\text{F}}(l_i)$ (e.g., model parameters, gradients, and optimizer states) and the variable memory components $\text{MEM}_{\text{V}}(l_i, p_i)$ (e.g., forward activations) during model training. 

\subsection{Profiling Acceleration}
\label{sec:Cost Estimation with Profiling}
To execute the 3D parallelism planning algorithm, AutoHet requires the computation time for each pipeline stage and peak memory to avoid out-of-memory (OOM) errors. We propose profiling acceleration strategies to expedite training initiation.

\noindent\textbf{\textit{Runtime profiling. }}The computation time for each pipeline stage is influenced by factors such as GPU types, workload distribution, and TP dimensions, leading to substantial profiling overhead. However, for repetitive architectures (e.g., GPT-2, LLaMA), we observe that the runtime of multiple model layers can be approximated by the cumulative runtime of individual layers with negligible error. Therefore, we adopt a binary decomposition approach, profiling iteration times for layer counts that are powers of two (e.g., 1, 2, 4, 8). Arbitrary layer counts are then represented as sums of these pre-profiled powers of two, as expressed in the following equation:
\begin{align}
T_{gpu}^{tp}(n) &= \sum_{i=0}^{k}\nolimits \alpha_{i} \cdot T_{gpu}^{tp}(2^{i}) 
\end{align}
where $T_{gpu}^{tp}(n)$ is the estimated iteration time for $n$ layers, $\alpha_{i}$ is a coefficient indicating the presence of the layer block in the decomposition of $n$, and $k = \lfloor \log_{2}{n} \rfloor$.

\noindent\textbf{\textit{Memory profiling. }}To reduce overhead, we introduce pruning strategies based on two observations: i), in repetitive model structures, memory consumption is mainly determined by the number of layers, with minimal dependence on their starting or ending points; ii), peak memory usage scales predictably with layer count. Consequently, we profile memory usage for a single model layer across different TP dimensions and derive the memory usage for multiple layers by summing the memory usage of individual layers.

\begin{algorithm}[!t]
\SetAlgoLined
\SetKwInOut{Input}{Input}
\SetKwInOut{Output}{Output}

\caption{3D Parallel Planning Algorithm}
\label{alg:3D Parallel Planning Algorithm}
\Input{Node specification $S$; Model config $\xi$;}
\Output{Optimal 3D parallel plan $P^{\ast}$ }

\begin{small}
\tcc{Initialize the valid TP dimensions}
\end{small}
$\mathrm{(t_{1} , t_{2}, \ldots, t_{n})} \gets getValidTpSize(U)$ \\

M $\gets estimateMemory(\xi)$;  X $\gets \left [\ \right]$; $T^{\ast } \gets \infty$ \\

\For{$tp\_dim \in (t_{1}, t_{2}, \ldots, t_{n})$}{
    \begin{small}
    \tcc{Modeling device grouping}
    \end{small}
    status, plan $\gets groupingDevice(U, M, \xi)$ \\
    \If{status}{
         Plans $\gets append(plan)$ \\}
}

\For{$plan \in Plans$}{
stage, node $\gets mapNodeAndStage(U,plan)$ \\
\begin{small}
\tcc{Modeling stages load balancing}
\end{small}
$\mathrm{P}$ $\gets balanceWorkload(stage,node,plan,\xi )$ \\
\If{$Cost(P) < T^{\ast}$}{
    $T^{\ast} = Cost(P)$  \\
    $P^{\ast}$ = $\mathrm{P}$ \\
    }
}
\Return $P^{\ast}$
\end{algorithm}

To summarize, Algorithm~\ref{alg:3D Parallel Planning Algorithm} demonstrates \system's 3D parallel planning algorithm for identifying better parallelism plans. It consists of three parts: 1) grouping devices to achieve a balanced computing power (Lines 3-8); 2) mapping GPU to nodes and stages to optimize communication efficiency (Line 10); and 3) balancing the workload across PP stages (Line 12). After generating several promising plans, AutoHet estimates the iteration times and select the optimal parallelism plan (Lines 14-16).

\section{Elastic Training Recovery}
\label{sec:elastictraining}

\subsection{Challenges of Training Recovery}

The availability of spot instances is time varying, depending on the demand of high-priority tasks. When training with GPU spot instances, a GPU currently in use can be preempted at the next billing slot, or a set of GPUs become available, which causes the updating of the parallelization plan. The reconfiguration overhead needs to be minimal in order to maintain continuity and efficiency in the parallel training process. Existing training frameworks such as DeepSpeed and Megatron-LM can use checkpointing strategies to handle spot instance preemption. Though feasible, these strategies are inefficient since the checkpoint is stored as a file at the GPU granularity, and transmitted to the cloud storage server. Consider a Llama-2 13B model, the checkpoint contains the full-precision optimizer state and the half-precision model weight, totaling 180GB in size. Thus, the uploading time to the cloud server and the downloading to all the participated GPUs is throttled by the communication bottleneck between the storage server and the training nodes.

The communication bandwidth between training nodes is typically an order of magnitude higher than that between a training node and the storage server. Therefore, a natural strategy is to retrieve training states from local nodes whenever possible and only fetch the missing pieces from the cloud storage server. This approach resembles failure-induced checkpoint recovery \cite{gimini,gandhi2024recycle}, but with key differences. In the prior literature, the entire checkpoint file of a GPU is replicated to a recovery node, whereas \system adopts a more fine-grained strategy: it generates checkpoint files at the layer level, tracks the physical locations of model partitions after each update to the 3D parallelization plan, and re-partitions the checkpoint to align with the new parallelization configuration.

\subsection{Design of Training Recovery}

\subsubsection{Layer-wise Checkpoint Generation}
 To handle dynamically changing parallelization plan, \system adopts a layer-wise checkpoint generation method and a hierarchical checkpoint storage scheme. In PyTorch, \textit{state\_dict} is a dictionary object used to save all crucial information during training including model parameter values as well as optimizer states such as momentum and variance. In general, \textit{state\_dict} is saved directly using the torch.save function and loaded during recovery, while it is no longer in a symmetric and standard shape in distributed training with heterogeneous types of GPUs. \system adopts a layer-wsie checkpoint generation procedure since a layer is the minimum unit of LLMs under different parallelization plans.

In \system, we filter the parameters of each layer by traversing all the key-value pairs of \textit{state\_dict}. Once each layer’s parameters are identified, they are extracted from the original \textit{state\_dict} and stored into a new dictionary object called \textit{layer\_dict}. \textit{layer\_dict} is organized by layers and contains only the parameters related to specific model layers, allowing for precise loading during recovery. Similarly, optimizer states are also stored at the layer granularity. \system creates an \textit{optimizer\_dict} that records the optimizer state for each layer including its momentum and variance, and stores them as independent dictionary entries. After filtering and reorganizing both model parameters and optimizer states, \system periodically saves the \textit{layer\_dict} and \textit{optimizer\_dict}  to CPU memory and disk using PyTorch’s torch.save function, and replicates them to the cloud storage server. Note that checkpoints should not be stored solely in CPU memory due to its volatile nature. In Kubernetes-managed environments, memory is cleared when processes are preempted or containers are rescheduled. Host machine SSDs provide persistent storage, ensuring data continuity and enabling recovery in such scenarios.

\subsubsection{Adaptive Checkpoint Loading}

\begin{figure*}[htbp]
\centering
\begin{subfigure}[b]{0.32\linewidth}
\centering
\includegraphics[width=\linewidth,height=3.5cm]{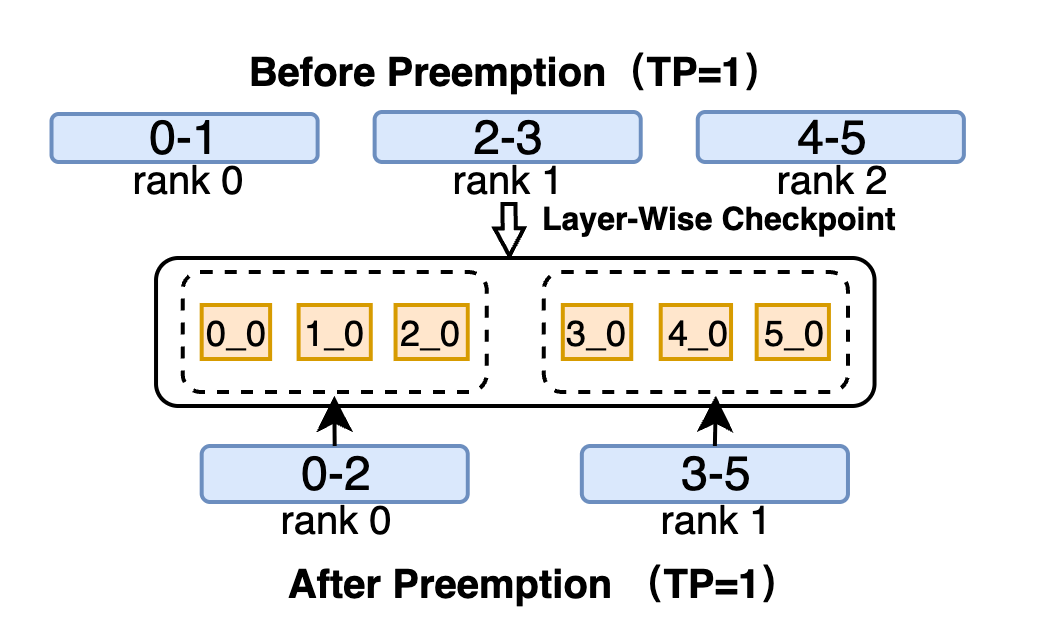}
\caption{TP dimension unaltered}
\label{fig:tp-nochange}
\end{subfigure}
\begin{subfigure}[b]{0.32\linewidth}
\centering
\includegraphics[width=\linewidth,height=3.5cm]{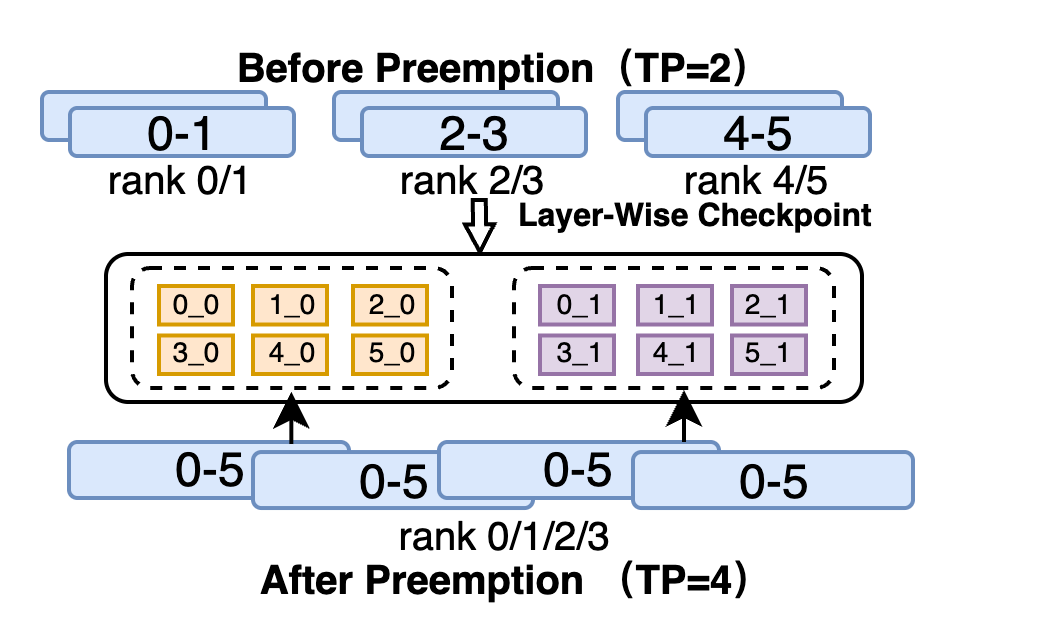}
\caption{TP dimension increased}
\label{fig:tp-increase}
\end{subfigure}
\begin{subfigure}[b]{0.32\linewidth}
\centering
\includegraphics[width=\linewidth,height=3.5cm]{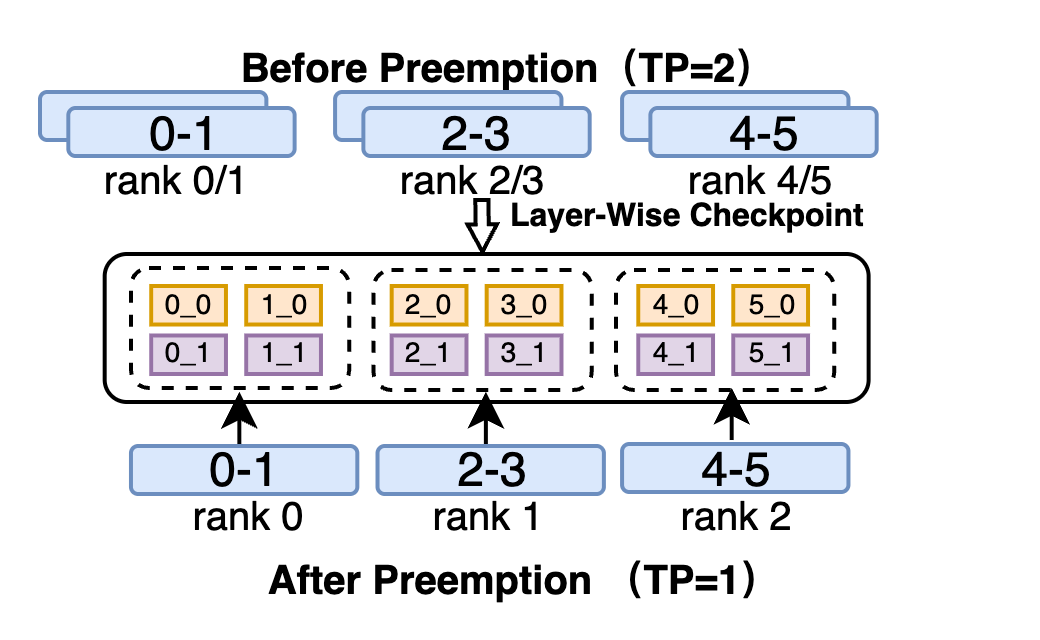}
\caption{TP dimension decreased}
\label{fig:tp-decrease}
\end{subfigure}
\vspace{-0.1cm}
\caption{Elastic Recovery Strategies for Parallelism Transitions}
\label{fig:一致分布吞吐量}
\vspace{-0.5cm}
\end{figure*}

After the parallelization plan changes, the training states need to be redistributed on the new set of GPUs. We design a parameter-level checkpoint loading scheme to flexibly adapt to this change. Specifically, it handles the following three scenarios: 

i) \emph{Unaltered TP dimension.} 
As shown in Figure \ref{fig:tp-nochange}, we suppose all GPUs are of the same model and the tensor parallelism (TP) dimension remains fixed at 1 for simplicity. Before preemption, rank 0–2 GPUs are pipeline-parallel, evenly splitting six model layers, with each GPU handling two layers of parameters. When rank 2 is preempted, rank 0 and rank 1 take over three layers each. Since the parameter partitioning has not changed, each current rank only needs to read the checkpoint files corresponding to its model layer numbers and TP rank ID. For example, rank 0 will read checkpoints $0\_0$, $1\_0$, and $2\_0$ (the first digit for rank ID and the second digit for TP partition ID), corresponding to TP rank 0 parameters for layers 0 to 2.

ii) \emph{Increased TP dimension.} In this case, directly loading layer-level checkpoints is not feasible, as their formats are not aligned and they must be re-partitioned according to the new TP dimension. In Figure \ref{fig:tp-increase}, after the preemption of ranks 4 and 5, the TP dimension increases from 2 to 4. Taking the loading of layer 0 as an example, the full parameters of layer 0 were initially stored in checkpoints $0\_0$ and $0\_1$. After the updating of the parallelization plan, the TP ranks expand to GPUs 0, 1, 2, and 3. Rank 0 and 1 need to read checkpoint $0\_0$, while ranks 2 and 3 read checkpoint $0\_1$. \system performs a split operation on each parameter matrix along the corresponding dimension in the checkpoint files, slices them into smaller blocks, and reorganizes them into a complete \textit{state\_dict} for each rank to load its required parameters.

iii) \emph{Decreased TP dimension.}
As shown in Figure \ref{fig:tp-decrease}, when ranks 3, 4, and 5 are preempted, the TP dimension decreases from 2 to 1. Again taking the loading of layer 0 as an example, after preemption, TP rank reduces to GPU 0. Here, TP rank 0 needs to read both checkpoints $0\_0$ and $0\_1$. \system performs the concatenation operation on each parameter matrix from the two checkpoint files along the corresponding dimension, merges them into complete parameter blocks, and reorganizes them into a complete \textit{state\_dict}.

\subsection{Accelerated Recovery}

\system designs an accelerated recovery strategy by leveraging a layer bitmap to record the physical locations of layer-wise checkpoints. Notably, a GPU spot instance may be reclaimed before the latest checkpoints are written to local storage, making them unavailable on the training nodes and only accessible from the cloud storage server. Hence, we consider two checkpoint transmission scenarios as follows. 

The first scenario occurs when, after resource changes, the checkpoints stored locally cannot be combined to form the complete model parameters and optimizer states. According to the layer bitmap, \system identifies which parameter layers each GPU needs and determines whether those parameters are already stored locally on that GPU. \system prioritizes to load those checkpoints stored at the local disk or CPU memory, and only fetches the missing pieces from the remote cloud. 

The second scenario is that the complete model parameters and optimizer states can be retrieved for the local training nodes. \system utilizes the RDMA links between training nodes to redistribute the training states, instead of downloading the checkpoints from the cloud storage server.

\section{Evaluation}
\label{sec:evaluation}

We implement \system in Python with 4000+ LOC (3165 for 3D parallelism and 1204 for elastic recovery), and our system modeules are integrated into Megatron-LM. Due to limited space, we do not elaborate the detailed implementation, but will open-source it later on.

\noindent{\textbf{\textit{Experimental setup.}}} We conduct our experiments using three types of NVIDIA GPUs: (i) A100 with 80GB HBM, (ii) H800 with 80GB HBM, and (iii) H20 with 100GB HBM. Our platform consists of four nodes, each equipped with eight GPUs. Specifically, Node 0 and Node 3 are A100 nodes, Node 1 is an H800 node, and Node 2 is an H20 node. Intra-node GPU communication is facilitated by NVLink, while inter-node communication uses 400Gbps RoCEv2. We evaluate three widely adopted LLMs including BERT-Large, GPT-3, and LLaMA, all of which utilize the Transformer architecture. Note that Node 3 is used exclusively in the evaluation of elastic recovery strategies.

\subsection{End-to-End Parallelization Performance}
We first compare the end-to-end training performance (tokens/s) of \system against two SOTA training systems: Megatron-LM and Whale. Since both systems lack automatic parallelism plans, we report their best-performing results m the parallel structures supported by each system. To evaluate \system's effectiveness, we cosnider two settings: a uniform GPU distribution, where each node is allocated with an equal number of GPUs; a non-uniform distribution, catering for real-world heterogeneous GPU environments, where the uniform GPU provisioning is often impractical. 

\begin{figure*}[htb] 
    \centering
    \begin{subfigure}{0.49\linewidth}
        \centering
        \includegraphics[width=\linewidth]{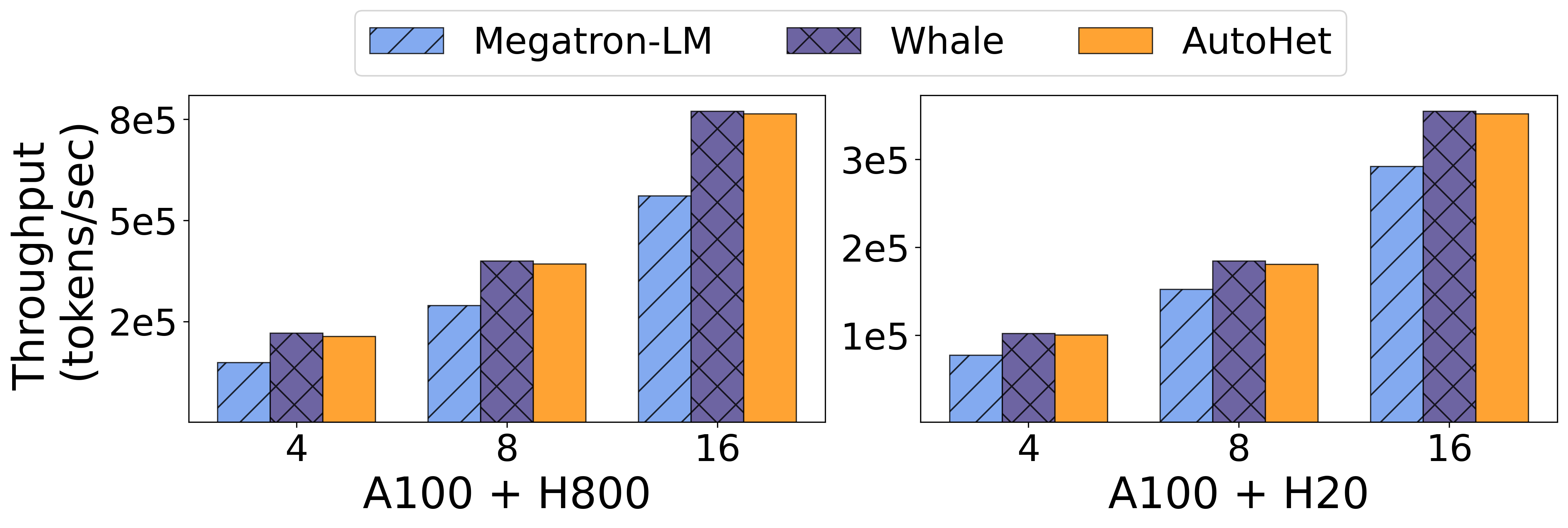}
        \caption{BERT-Large}
        \label{fig:uniform-bert}
    \end{subfigure}
    \hfill 
    \begin{subfigure}{0.49\linewidth}
        \centering
        \includegraphics[width=\linewidth]{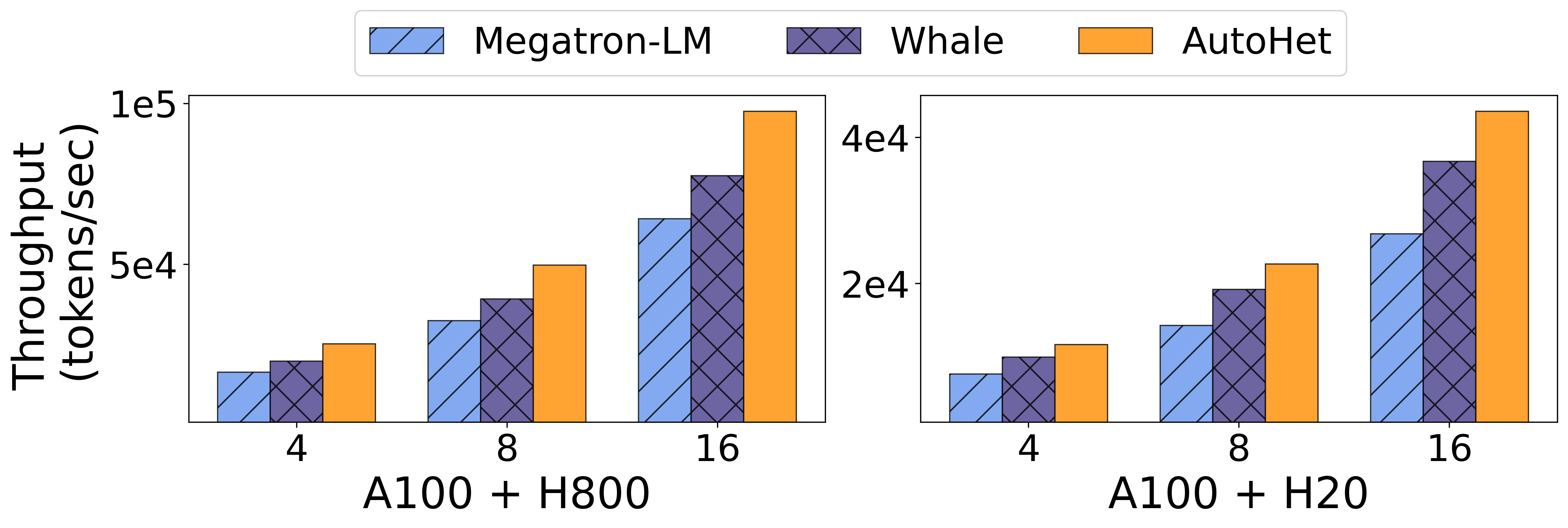} 
        \caption{GPT-3}
        \label{fig:uniform-gpt}
    \end{subfigure}
    \vspace{-0.2cm}
    \caption{End-to-end performance evaluation under a uniform GPU distribution for differing GPU types and numbers.}
    \vspace{-0.5cm}
    \label{fig:uniform-eva}
\end{figure*}

\begin{figure*}[t] 
    \centering
    \begin{subfigure}{0.48\linewidth}
        \centering
        \includegraphics[width=0.95\linewidth]{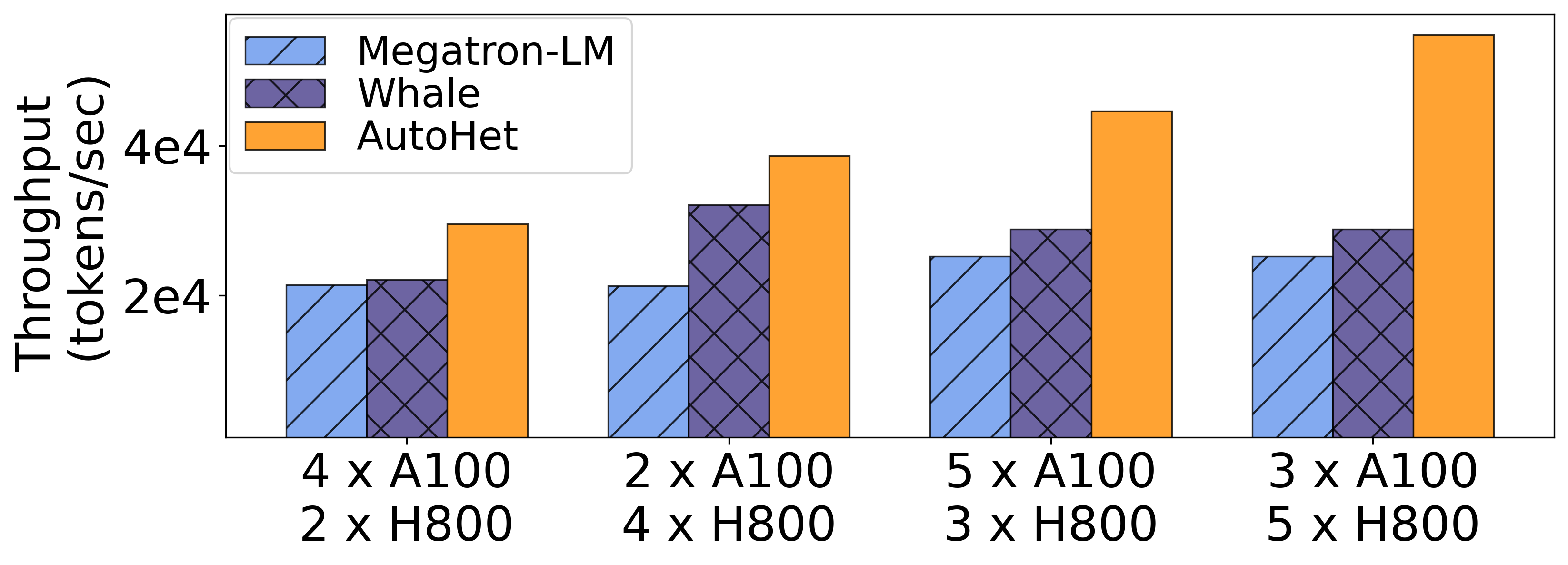}
        \caption{A100 + H800}
    \end{subfigure}
    \hfill 
    \begin{subfigure}{0.48\linewidth}
        \centering
        \includegraphics[width=0.95\linewidth]{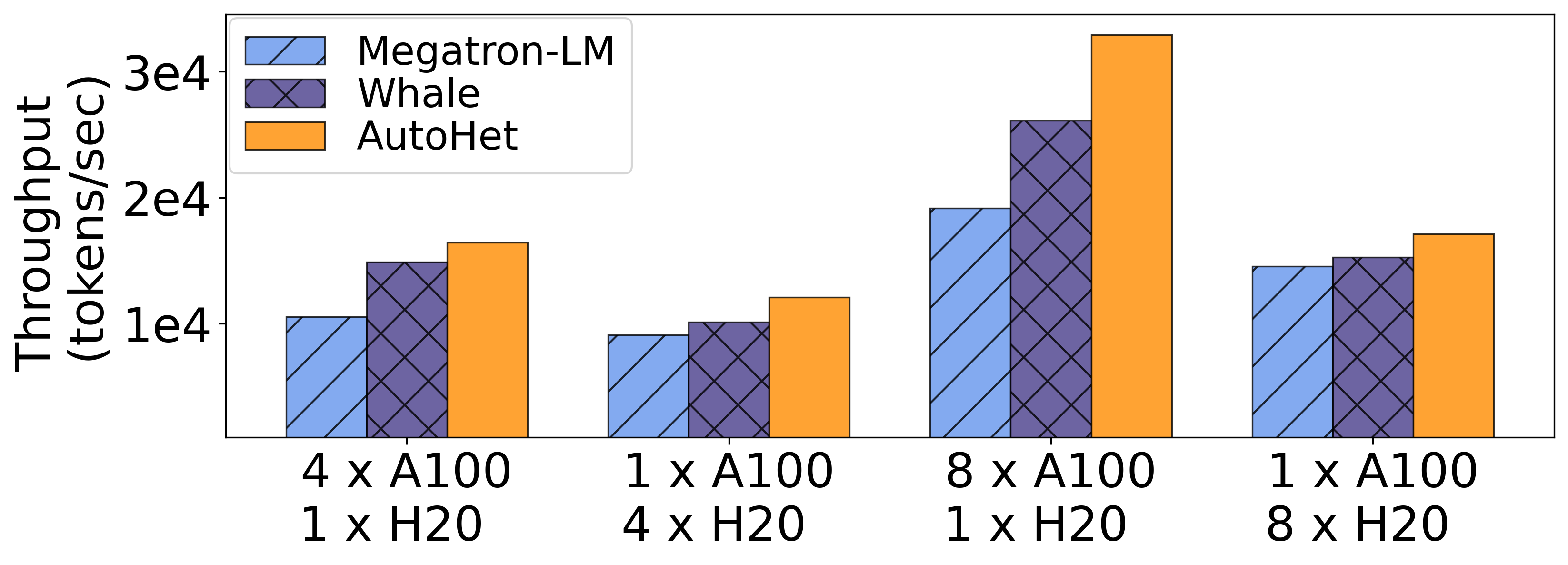} 
        \caption{A100 + H20}
    \end{subfigure}
    \vspace{-0.2cm}
    \caption{End-to-end performance evaluation under a non-uniform GPU distribution for differing GPU types and numbers.}
    \vspace{-0.5cm}
    \label{fig:nonuniform}
\end{figure*}

\subsubsection{Uniform GPU Distribution} 
We evaluate two GPU type combinations: H800+A100 and A100+H20, with each node configured with 2, 4, or 8 GPUs. We select two model architectures: BERT-Large with 340M parameters and GPT-3 with 6.7B parameters. The training performance is displayed in Figure~\ref{fig:uniform-eva}. 

\noindent{\textbf{\textit{BERT-Large results. }}} AutoHet achieves an average training throughput 1.38$\times$ higher than Megatron-LM across all experiments. Due to the relatively small size of BERT-Large, it can be stored in a single GPU of any type. Hence, Megatron-LM directly adopts the full data parallelism despite GPU heterogeneity, which causes the severe straggler problem. Whale readjusts the batch size on each GPU according to its computing power (referred to as ``Intra-TaskGraph load balance"), enabling it to achieve effective performance under these conditions. AutoHet employs a hybrid approach that combines PP and DP, while implementing load balancing across different pipeline stages. Nevertheless, AutoHet automatically generates parallelism plan and attains comparable performance without adversely affecting convergence.

\noindent{\textbf{\textit{GPT-3 results. }}}Across all experiments, AutoHet outperforms Megatron-LM and Whale on average 1.53$\times$ and 1.27$\times$ in training throughput. Performance improvement can be attributed to two major factors: i), unlike BERT-Large, GPT-3 cannot be stored on a single GPU due to its relatively large model size, necessitating model parallelism. Workload balancing method in AutoHet addresses this by efficiently distributing model layers across heterogeneous GPUs. In contrast, Megatron-LM assumes homogeneous GPUs, resulting in a uniform layer division that fails to exploit the performance variations across different GPU types; ii), AutoHet schedules GPUs with lower computing power to earlier pipeline stages, enabling GPUs with higher computational power to handle larger portions of the workload. This is in contrast to Megatron-LM and Whale, which allocate stages based on a sequential GPU node order without considering the specific performance characteristics of each node.

\begin{figure}[t]
    \centering
    \includegraphics[width=0.96\linewidth]{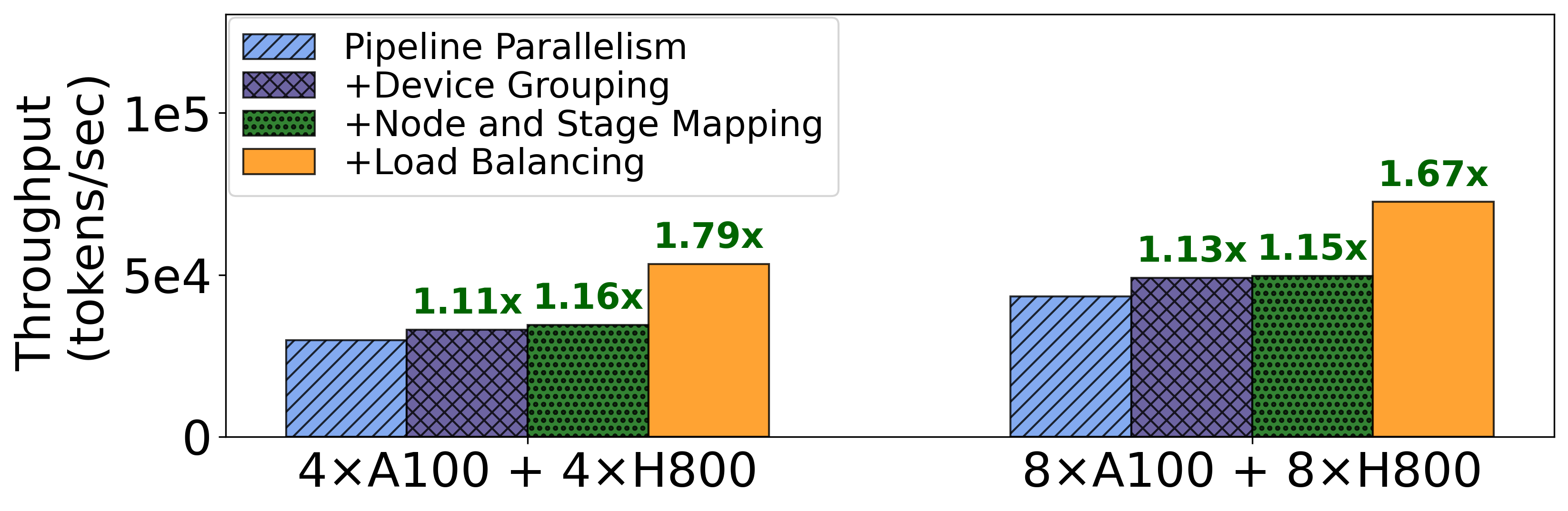}
    \caption{Performance Breakdown of the GPT-3 6.7B Model}
    \vspace{-0.5cm}
    \label{fig:breakdown}
\end{figure}

\begin{figure*}[t]
    \centering
    \begin{subfigure}{0.32\linewidth}
        \centering
        \includegraphics[width=1\linewidth]{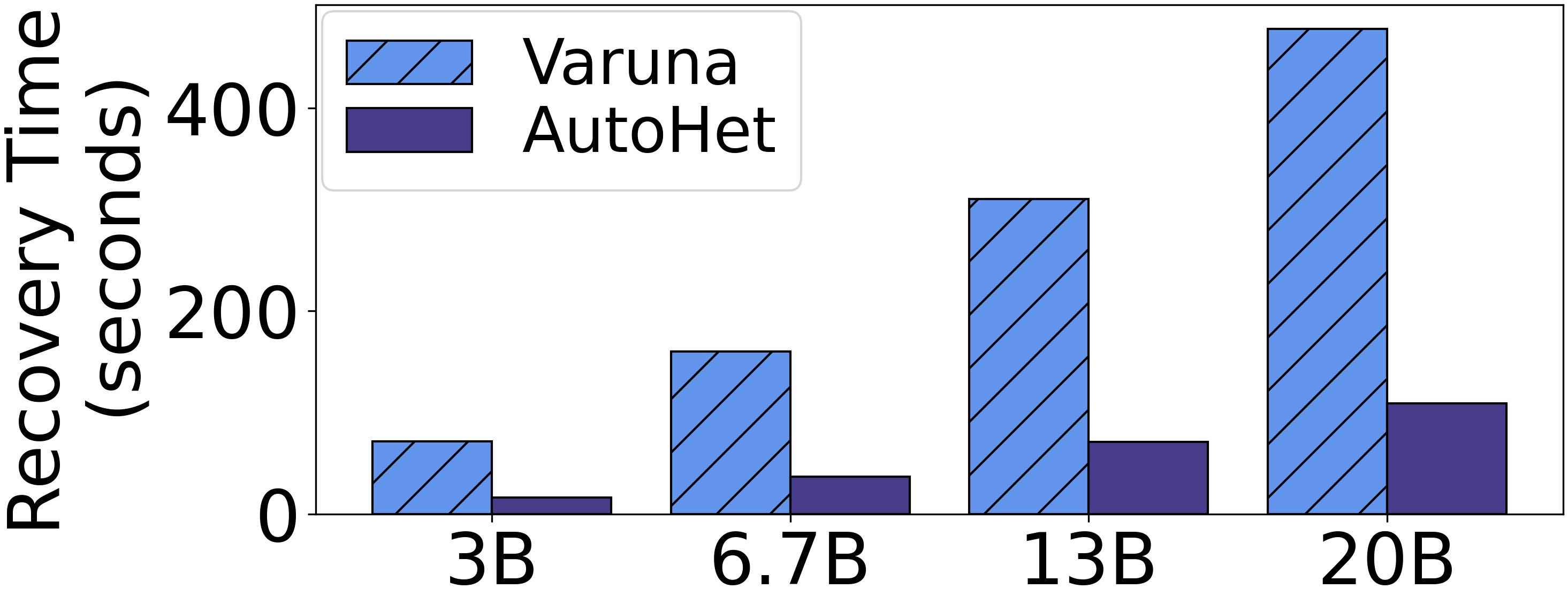}
        \caption{Local Storage-based Recovery}
        \label{fig:speed1}
    \end{subfigure}
    \hfill 
    \begin{subfigure}{0.32\linewidth}
        \centering
        \includegraphics[width=1\linewidth]{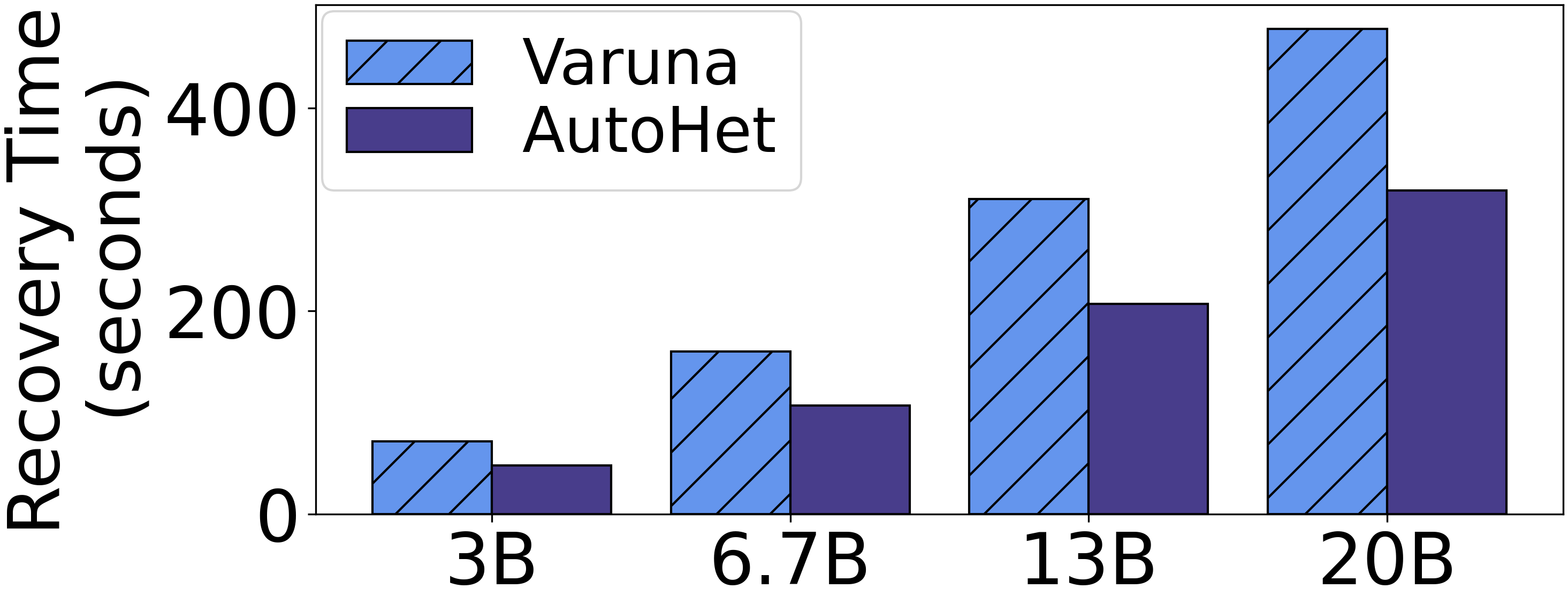}
        \caption{Hybrid Local-Cloud Recovery}
        \label{fig:speed2}
    \end{subfigure}
    \hfill
    \begin{subfigure}{0.32\linewidth}
        \centering
        \includegraphics[width=1\linewidth]{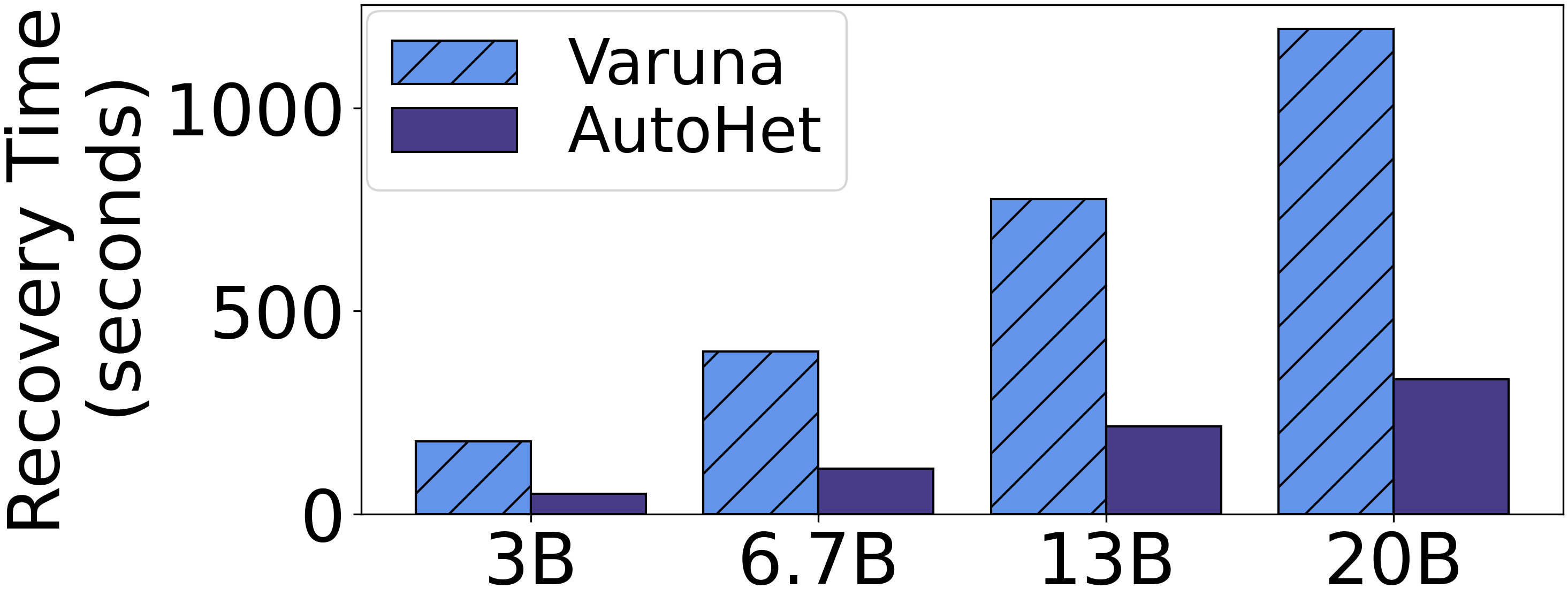}
        \caption{RDMA-accelerated Recovery}
        \label{fig:speed3}
    \end{subfigure}
    \vspace{-0.2cm}
    \caption{Elastic Recovery Performance Analysis under Different Bandwidth Utilization Strategies in AutoHet}
    \vspace{-0.5cm}
    \label{fig:recoveryspeed-eva}
\end{figure*}

\subsubsection{Non-uniform GPU Distribution}
We select the LLaMA model with 6.7B parameters and evaluate two distinct GPU type combinations: H800+A100 and A100+H20. Figure~\ref{fig:nonuniform} illustrates the training performance across various combinations of GPU quantities.

For the H800+A100 combination, AutoHet significantly outperforms Megatron-LM and Whale by average factors of 1.79$\times$ and 1.51$\times$ in training throughput, respectively. The main reason for this improvement is AutoHet's support for asymmetric parallel structures, whereas the baseline systems are constrained by symmetric parallelism. For instance, in the 4$\times$A100+2$\times$H800 experimental setting, AutoHet can configure a TP degree of two, with H800 models forming individual DP groups and A100 models constituting another DP group with a two-stage pipeline. In contrast, Megatron-LM and Whale are limited to forming costly long pipeline parallel structures. Similarly, in configurations such as 5$\times$A100+3$\times$H800 and 3$\times$A100+5$\times$H800, when the number of GPUs per node is odd (preventing the formation of TP groups), AutoHet can create a more balanced combination of pipeline and data parallelism, while Megatron-LM and Whale are unable to achieve load balancing across model layers due to their inability to support an inconsistent number of layers within the same stage across different DP groups.

For the A100+H20 combination, we design experimental settings with larger quantity disparities between GPU types. Across all configurations, Megatron-LM and Whale are constrained to adopt pure pipeline parallelism approaches, as they cannot support structures with inconsistent GPU counts across different data parallel groups. In contrast, AutoHet demonstrates remarkable flexibility in its parallel structures. For example, in the 1$\times$A100+4$\times$H20 experimental setting, AutoHet can form two DP groups: one comprising 1$\times$A100+1$\times$H20, and another consisting of 3$\times$H20, with pipeline parallelism employed within each DP group. This adaptive parallelization strategy enables AutoHet to achieve average speedups of 1.44$\times$ and 1.16$\times$ over Megatron-LM and Whale respectively.

\subsection{Breakdown Analysis and System Overheads}
We conduct the performance breakdown experiments on two heterogeneous GPU configurations: 4$\times$A100+4$\times$H800 and 8$\times$A100+8$\times$H800, as shown in Figure~\ref{fig:breakdown}. Since both experimental settings demonstrate similar improvement trends, we analyze the results m the 4$\times$A100 + 4$\times$H800 setup to evaluate the contribution of each component of AutoHet. Using GPT-3 6.7B as a case study and basic pipeline parallelism training as the baseline, we measure the cumulative benefit of incrementally adding each module compared to the baseline. We observe that the device grouping module achieves a 1.11$\times$ increase in throughput, and this gain stems m the reduced bubble ratio in pipeline parallelism. The node and stage mapping module further improves the throughput gain to 1.16$\times$ higher than the baseline. The workload balancing between stages contributes a 1.79$\times$ throughput gain, in which the transformer layers are appropriately allocated to the GPUs of heterogeneous computing powers along the pipelines. 

The profiling and planning overheads are crucial to fairly evaluate the performance of \system, especially in the context of spot instance training. We directly use SCIP \cite{scip} to solve the non-linear integer programming problem in \system. When the number of GPUs are in the set $\{16, 24, 32, 64\}$, the planning times for search the optimal parallelization strategies are $\{1.23, 5.72, 16.96, 159.12\}$ seconds, which are acceptable to practical spot instance training scenarios. In contrast, Alpa~\cite{zheng2022alpa} demands 240 minutes to search for the intra- and inter-operator parallelization strategy with homogeneous GPUs, as reported in~\cite{um2024metis}. The reason of such improvement stems m that a huge amount of infeasible schemes have been eliminated in the mathematical framework of \system based on our observations and domain specific simplifications. We further measure the realistic runtime of each layer, and emulate the profiling time overheads using emulations. With the same set of GPUs, the profiling time of \system increases m 11.9 to 15.4 minutes, and this overhead does not greatly scale up with the number of GPUs or the model size. 
Compared with Alpa that requires the profiling time of 209 minutes~\cite{um2024metis}, \system is nearly ten times faster. 

\subsection{Elastic Recovery Time}

\emph{Recovery time} is the metric that measures the time required for training to continue after preemption and recovery. We hereby consider different resource configurations, parallel strategies and model scales to comprehensively evaluate the \system's recovery efficiency. Our baseline strategy, Varuna \cite{varuna}, is a low-cost elastic training system designed for spot instances. It supports a hybrid data and pipeline parallel training paradigm, and enables the hierarchical checkpoint storage and loading. Since Varuna does not support checkpoint recovery under tensor parallelism, we only compare its checkpoint fetching strategy with \system. The experiments were conducted on GPT-3 models of sizes 3B, 6.7B, 13B, and 20B, with cloud bandwidth set to 1200MB/s~\cite{alibaba_extreme_nas_2025} and local storage utilizing NVMe SSDs achieving 3500MB/s end-to-end checkpoint loading bandwidth.

Figure \ref{fig:recoveryspeed-eva} shows the recovery times of Varuna and \system under three different scenarios. In scenario A, node $N_0$ has 8 A100 GPUs and node $N_1$ has 8 H20 GPUs. The current 3D parallelization strategy consists of four DP groups with each group uses 2 A100 GPUs and 2 H20 GPUs for PP and TP. When two DP groups (4 A100 and 4 H20 GPUs) are completely preempted, the local nodes maintain complete checkpoint replicas, enabling direct local access to all required training states. The training of Varuna pauses and downloads the checkpoint m the cloud storage for recovery, while \system simply loads the checkpoint locally, significantly reducing the recovery time and achieving a 4.38$\times$ speedup. In scenario B, the eight GPUs of node 0 are preempted so that the original parallelization strategy is changed to two DP groups, each group containing 4 H20 GPUs operated in TP. Only the part of the checkpoint is available locally according to the constructed layer bitmap, and thus missing part is retrieved m the cloud, achieving a 1.49$\times$ speedup compared with Varuna. The scenario C emulates the increase of avaialable spot GPU instances. We augment two nodes, $N_2$ with 2 A100 GPUs and $N_3$ with 2 H20 GPUs. The new parallelization strategy contains one more DP group, and the training state can be obtained entirely m local machines through RDMA links. In contrast, the cloud-based retrieval strategy becomes increasingly inefficient as the number of DP groups scales up, requiring the download of larger volumes of complete model parameters. This scalability limitation further demonstrates the superiority of AutoHet's  accelerated recovery approach which is 3.59$\times$ faster than Varuna. 
\section{Conclusion}
We present AutoHet, an automated parallelization framework designed for distributed training on heterogeneous spot instance GPU clusters. By addressing the challenges of implementing 3D parallelism in diverse GPU environments, AutoHet supports asymmetric parallel structures and employs a novel 3D parallel planning algorithm to optimize workload distribution and minimize overheads. Our comprehensive experiments across three distinct GPU types and three different model architectures demonstrate that AutoHet achieves up to 1.79× throughput improvement over existing systems, and delivers 1.49× to 4.38× speedups in elastic recovery time.



\bibliographystyle{IEEEtran}
\bibliography{sample-base}

@article{gpt3,
  title={Language models are few-shot learners},
  author={Brown, Tom and Mann, Benjamin and Ryder, Nick and Subbiah, Melanie and Kaplan, Jared D and Dhariwal, Prafulla and Neelakantan, Arvind and Shyam, Pranav and Sastry, Girish and Askell, Amanda and others},
  journal={Advances in neural information processing systems},
  volume={33},
  pages={1877--1901},
  year={2020}
}

@article{touvron2023llama,
  title={Llama: Open and efficient foundation language models},
  author={Touvron, Hugo and Lavril, Thibaut and Izacard, Gautier and Martinet, Xavier and Lachaux, Marie-Anne and Lacroix, Timoth{\'e}e and Rozi{\`e}re, Baptiste and Goyal, Naman and Hambro, Eric and Azhar, Faisal and others},
  journal={arXiv preprint arXiv:2302.13971},
  year={2023}
}

@article{team2024gemini,
  title={Gemini 1.5: Unlocking multimodal understanding across millions of tokens of context},
  author={Team, Gemini and Georgiev, Petko and Lei, Ving Ian and Burnell, Ryan and Bai, Libin and Gulati, Anmol and Tanzer, Garrett and Vincent, Damien and Pan, Zhufeng and Wang, Shibo and others},
  journal={arXiv preprint arXiv:2403.05530},
  year={2024}
}

@inproceedings{hetecluster,
  title={Analysis of $\{$Large-Scale$\}$$\{$Multi-Tenant$\}$$\{$GPU$\}$ clusters for $\{$DNN$\}$ training workloads},
  author={Jeon, Myeongjae and Venkataraman, Shivaram and Phanishayee, Amar and Qian, Junjie and Xiao, Wencong and Yang, Fan},
  booktitle={2019 USENIX Annual Technical Conference (USENIX ATC 19)},
  pages={947--960},
  year={2019}
}

@inproceedings{weng2022mlaas,
  title={$\{$MLaaS$\}$ in the wild: Workload analysis and scheduling in $\{$Large-Scale$\}$ heterogeneous $\{$GPU$\}$ clusters},
  author={Weng, Qizhen and Xiao, Wencong and Yu, Yinghao and Wang, Wei and Wang, Cheng and He, Jian and Li, Yong and Zhang, Liping and Lin, Wei and Ding, Yu},
  booktitle={19th USENIX Symposium on Networked Systems Design and Implementation (NSDI 22)},
  pages={945--960},
  year={2022}
}

@misc{nvidia_blackwell_architecture,
  author       = {NVIDIA},
  title        = {Blackwell Architecture},
  howpublished = {\url{https://www.nvidia.com/en-us/data-center/technologies/blackwell-architecture/}},
  year         = {2023},
  note         = {2025-01-08}
}

@article{2019megatron,
  title={Megatron-lm: Training multi-billion parameter language models using model parallelism},
  author={Shoeybi, Mohammad and Patwary, Mostofa and Puri, Raul and LeGresley, Patrick and Casper, Jared and Catanzaro, Bryan},
  journal={arXiv preprint arXiv:1909.08053},
  year={2019}
}

@inproceedings{2021megatron,
  title={Efficient large-scale language model training on gpu clusters using megatron-lm},
  author={Narayanan, Deepak and Shoeybi, Mohammad and Casper, Jared and LeGresley, Patrick and Patwary, Mostofa and Korthikanti, Vijay and Vainbrand, Dmitri and Kashinkunti, Prethvi and Bernauer, Julie and Catanzaro, Bryan and others},
  booktitle={Proceedings of the International Conference for High Performance Computing, Networking, Storage and Analysis},
  pages={1--15},
  year={2021}
}

@inproceedings{2020deepspeed,
  title={Deepspeed: System optimizations enable training deep learning models with over 100 billion parameters},
  author={Rasley, Jeff and Rajbhandari, Samyam and Ruwase, Olatunji and He, Yuxiong},
  booktitle={Proceedings of the 26th ACM SIGKDD International Conference on Knowledge Discovery \& Data Mining},
  pages={3505--3506},
  year={2020}
}

@inproceedings{2021tp,
  title={Model-parallel model selection for deep learning systems},
  author={Nagrecha, Kabir},
  booktitle={Proceedings of the 2021 international conference on management of data},
  pages={2929--2931},
  year={2021}
}

@article{huang2019gpipe,
  title={Gpipe: Efficient training of giant neural networks using pipeline parallelism},
  author={Huang, Yanping and Cheng, Youlong and Bapna, Ankur and Firat, Orhan and Chen, Dehao and Chen, Mia and Lee, HyoukJoong and Ngiam, Jiquan and Le, Quoc V and Wu, Yonghui and others},
  journal={Advances in neural information processing systems},
  volume={32},
  year={2019}
}

@inproceedings{narayanan2019pipedream,
  title={PipeDream: Generalized pipeline parallelism for DNN training},
  author={Narayanan, Deepak and Harlap, Aaron and Phanishayee, Amar and Seshadri, Vivek and Devanur, Nikhil R and Ganger, Gregory R and Gibbons, Phillip B and Zaharia, Matei},
  booktitle={Proceedings of the 27th ACM symposium on operating systems principles},
  pages={1--15},
  year={2019}
}

@inproceedings{pipedream2bw,
  title={Memory-efficient pipeline-parallel dnn training},
  author={Narayanan, Deepak and Phanishayee, Amar and Shi, Kaiyu and Chen, Xie and Zaharia, Matei},
  booktitle={International Conference on Machine Learning},
  pages={7937--7947},
  year={2021},
  organization={PMLR}
}

@inproceedings{rajbhandari2020zero,
  title={Zero: Memory optimizations toward training trillion parameter models},
  author={Rajbhandari, Samyam and Rasley, Jeff and Ruwase, Olatunji and He, Yuxiong},
  booktitle={SC20: International Conference for High Performance Computing, Networking, Storage and Analysis},
  pages={1--16},
  year={2020},
  organization={IEEE}
}

@article{scip,
  title={SCIP: solving constraint integer programs},
  author={Achterberg, Tobias},
  journal={Mathematical Programming Computation},
  volume={1},
  pages={1--41},
  year={2009},
  publisher={Springer}
}

@article{pytorch,
  title={Pytorch: An imperative style, high-performance deep learning library},
  author={Paszke, Adam and Gross, Sam and Massa, Francisco and Lerer, Adam and Bradbury, James and Chanan, Gregory and Killeen, Trevor and Lin, Zeming and Gimelshein, Natalia and Antiga, Luca and others},
  journal={Advances in neural information processing systems},
  volume={32},
  year={2019}
}

@misc{bekman2022bloom,
  author       = {Stas Bekman},
  title        = {The Technology Behind Bloom Training},
  year         = 2022,
  howpublished = {\url{https://huggingface.co/blog/bloom-megatron-deepspeed}}
}

@misc{nvidia_cudagpus,
  author       = {NVIDIA Developer},
  title        = {CUDA GPUs},
  year         = 2024,
  url          = {https://developer.nvidia.cn/cuda-gpus},
  note         = {Accessed: 2024-07-29}
}

@article{shazeer2018mesh,
  title={Mesh-tensorflow: Deep learning for supercomputers},
  author={Shazeer, Noam and Cheng, Youlong and Parmar, Niki and Tran, Dustin and Vaswani, Ashish and Koanantakool, Penporn and Hawkins, Peter and Lee, HyoukJoong and Hong, Mingsheng and Young, Cliff and others},
  journal={Advances in neural information processing systems},
  volume={31},
  year={2018}
}

@article{dean2012large,
  title={Large scale distributed deep networks},
  author={Dean, Jeffrey and Corrado, Greg and Monga, Rajat and Chen, Kai and Devin, Matthieu and Mao, Mark and Ranzato, Marc'aurelio and Senior, Andrew and Tucker, Paul and Yang, Ke and others},
  journal={Advances in neural information processing systems},
  volume={25},
  year={2012}
}

@online{microsoftazure,
    author       = {Microsoft},
    title        = {Microsoft Azure Pricing},
    year         = {2024},
    url          = {https://azure.microsoft.com/en-us/pricing/}
}

@online{amazonec2,
    author       = {Amazon},
    title        = {Amazon EC2 Pricing},
    year         = {2024},
    url          = {https://aws.amazon.com/ec2/pricing/}
}

@inproceedings{2019-bert,
    title = "{BERT}: Pre-training of Deep Bidirectional Transformers for Language Understanding",
    author = "Devlin, Jacob  and
      Chang, Ming-Wei  and
      Lee, Kenton  and
      Toutanova, Kristina",
    editor = "Burstein, Jill  and
      Doran, Christy  and
      Solorio, Thamar",
    booktitle = "Proceedings of the 2019 Conference of the North {A}merican Chapter of the Association for Computational Linguistics: Human Language Technologies, Volume 1 (Long and Short Papers)",
    month = jun,
    year = "2019",
    address = "Minneapolis, Minnesota",
    publisher = "Association for Computational Linguistics",
    url = "https://aclanthology.org/N19-1423",
    pages = "4171--4186",
}

@inproceedings{jiang2017heterogeneity,
  title={Heterogeneity-aware distributed parameter servers},
  author={Jiang, Jiawei and Cui, Bin and Zhang, Ce and Yu, Lele},
  booktitle={Proceedings of the 2017 ACM International Conference on Management of Data},
  pages={463--478},
  year={2017}
}

@inproceedings{jia2022whale,
  title={Whale: Efficient giant model training over heterogeneous $\{$GPUs$\}$},
  author={Jia, Xianyan and Jiang, Le and Wang, Ang and Xiao, Wencong and Shi, Ziji and Zhang, Jie and Li, Xinyuan and Chen, Langshi and Li, Yong and Zheng, Zhen and others},
  booktitle={2022 USENIX Annual Technical Conference (USENIX ATC 22)},
  pages={673--688},
  year={2022}
}

@inproceedings{zheng2022alpa,
  title={Alpa: Automating inter-and $\{$Intra-Operator$\}$ parallelism for distributed deep learning},
  author={Zheng, Lianmin and Li, Zhuohan and Zhang, Hao and Zhuang, Yonghao and Chen, Zhifeng and Huang, Yanping and Wang, Yida and Xu, Yuanzhong and Zhuo, Danyang and Xing, Eric P and others},
  booktitle={16th USENIX Symposium on Operating Systems Design and Implementation (OSDI 22)},
  pages={559--578},
  year={2022}
}

@article{zhou2023abs,
  title={ABS-SGD: A Delayed Synchronous Stochastic Gradient Descent Algorithm with Adaptive Batch Size for Heterogeneous GPU Clusters},
  author={Zhou, Xin and Chen, Ling and Wu, Houming},
  journal={arXiv preprint arXiv:2308.15164},
  year={2023}
}

@inproceedings{song2020accpar,
  title={Accpar: Tensor partitioning for heterogeneous deep learning accelerators},
  author={Song, Linghao and Chen, Fan and Zhuo, Youwei and Qian, Xuehai and Li, Hai and Chen, Yiran},
  booktitle={2020 IEEE International Symposium on High Performance Computer Architecture (HPCA)},
  pages={342--355},
  year={2020},
  organization={IEEE}
}

@inproceedings{chen2020semi,
  title={Semi-dynamic load balancing: Efficient distributed learning in non-dedicated environments},
  author={Chen, Chen and Weng, Qizhen and Wang, Wei and Li, Baochun and Li, Bo},
  booktitle={Proceedings of the 11th ACM Symposium on Cloud Computing},
  pages={431--446},
  year={2020}
}

@inproceedings{ding2021hetseq,
  title={HetSeq: distributed GPU training on heterogeneous infrastructure},
  author={Ding, Yifan and Botzer, Nicholas and Weninger, Tim},
  booktitle={Proceedings of the AAAI Conference on Artificial Intelligence},
  volume={35},
  number={17},
  pages={15432--15438},
  year={2021}
}

@inproceedings{park2020hetpipe,
  title={$\{$HetPipe$\}$: Enabling large $\{$DNN$\}$ training on (whimpy) heterogeneous $\{$GPU$\}$ clusters through integration of pipelined model parallelism and data parallelism},
  author={Park, Jay H and Yun, Gyeongchan and Chang, M Yi and Nguyen, Nguyen T and Lee, Seungmin and Choi, Jaesik and Noh, Sam H and Choi, Young-ri},
  booktitle={2020 USENIX Annual Technical Conference (USENIX ATC 20)},
  pages={307--321},
  year={2020}
}

@inproceedings{duan2022hph,
  title={Hph: Hybrid parallelism on heterogeneous clusters for accelerating large-scale dnns training},
  author={Duan, Yabo and Lai, Zhiquan and Li, Shengwei and Liu, Weijie and Ge, Keshi and Liang, Peng and Li, Dongsheng},
  booktitle={2022 IEEE International Conference on Cluster Computing (CLUSTER)},
  pages={313--323},
  year={2022},
  organization={IEEE}
}

@inproceedings{um2024metis,
  title={Metis: Fast Automatic Distributed Training on Heterogeneous $\{$GPUs$\}$},
  author={Um, Taegeon and Oh, Byungsoo and Kang, Minyoung and Lee, Woo-Yeon and Kim, Goeun and Kim, Dongseob and Kim, Youngtaek and Muzzammil, Mohd and Jeon, Myeongjae},
  booktitle={2024 USENIX Annual Technical Conference (USENIX ATC 24)},
  pages={563--578},
  year={2024}
}

@article{miao2023sdpipe,
  title={Sdpipe: A semi-decentralized framework for heterogeneity-aware pipeline-parallel training},
  author={Miao, Xupeng and Shi, Yining and Yang, Zhi and Cui, Bin and Jia, Zhihao},
  journal={Proceedings of the VLDB Endowment},
  volume={16},
  number={9},
  pages={2354--2363},
  year={2023},
  publisher={VLDB Endowment}
}

@inproceedings{varuna,
author = {Athlur, Sanjith and Saran, Nitika and Sivathanu, Muthian and Ramjee, Ramachandran and Kwatra, Nipun},
title = {Varuna: scalable, low-cost training of massive deep learning models},
year = {2022},
isbn = {9781450391627},
publisher = {Association for Computing Machinery},
address = {New York, NY, USA},
booktitle = {Proceedings of the Seventeenth European Conference on Computer Systems},
pages = {472–487},
numpages = {16},
location = {Rennes, France},
series = {EuroSys '22}
}

@inproceedings{gimini,
author = {Wang, Zhuang and Jia, Zhen and Zheng, Shuai and Zhang, Zhen and Fu, Xinwei and Ng, T. S. Eugene and Wang, Yida},
title = {GEMINI: Fast Failure Recovery in Distributed Training with In-Memory Checkpoints},
year = {2023},
isbn = {9798400702297},
publisher = {Association for Computing Machinery},
address = {New York, NY, USA},
booktitle = {Proceedings of the 29th Symposium on Operating Systems Principles},
pages = {364–381},
numpages = {18},
location = {Koblenz, Germany},
series = {SOSP '23}
}

@inproceedings{gandhi2024recycle,
  title={Recycle: Resilient training of large dnns using pipeline adaptation},
  author={Gandhi, Swapnil and Zhao, Mark and Skiadopoulos, Athinagoras and Kozyrakis, Christos},
  booktitle={Proceedings of the ACM SIGOPS 30th Symposium on Operating Systems Principles},
  pages={211--228},
  year={2024}
}

@misc{alibaba_extreme_nas_2025,
    title        = {Extreme {NAS} File Systems - Product Overview},
    author       = {{Alibaba Cloud}},
    organization = {Alibaba Cloud Computing Co., Ltd.},
    year         = {2025},
    url          = {https://help.aliyun.com/zh/nas/product-overview/extreme-nas-file-systems},
    urldate      = {2025-01-08},
    note         = {Online documentation}
}

\end{document}